 \documentstyle[epsfig]{mn}

\newif\ifAMStwofonts

\def \csim{$\sim$}
\def \ell{$\epsilon$}
\def \csig{$\sigma$}
\def \hth{$h_{3}$}
\def \hf{$h_{4}$}

\def \kms{\rm km~s$^{-1}$}

\ifoldfss
  \ifCUPmtlplainloaded \else
    \NewTextAlphabet{textbfit} {cmbxti10} {}
    \NewTextAlphabet{textbfss} {cmssbx10} {}
    \NewMathAlphabet{mathbfit} {cmbxti10} {} 
    \NewMathAlphabet{mathbfss} {cmssbx10} {} 
  \fi
  \ifAMStwofonts
    \ifCUPmtlplainloaded \else
      \NewSymbolFont{upmath} {eurm10}
      \NewSymbolFont{AMSa} {msam10}
      \NewMathSymbol{\upi}     {0}{upmath}{19}
      \NewMathSymbol{\umu}     {0}{upmath}{16}
      \NewMathSymbol{\upartial}{0}{upmath}{40}
      \NewMathSymbol{\leqslant}{3}{AMSa}{36}
      \NewMathSymbol{\geqslant}{3}{AMSa}{3E}

      \let\leq=\leqslant \let\le=\leqslant
       \let\ge=\geqslant
    \fi
  \fi
\fi 

\ifnfssone
  \newmathalphabet{\mathit}
  \addtoversion{normal}{\mathit}{cmr}{m}{it}
  \addtoversion{bold}{\mathit}{cmr}{bx}{it}
  \newmathalphabet{\mathbfit} 
  \addtoversion{normal}{\mathbfit}{cmr}{bx}{it}
  \addtoversion{bold}{\mathbfit}{cmr}{bx}{it}
  \newmathalphabet{\mathbfss} 
  \addtoversion{normal}{\mathbfss}{cmss}{bx}{n}
  \addtoversion{bold}{\mathbfss}{cmss}{bx}{n}
  \ifAMStwofonts
    \ifCUPmtlplainloaded \else
      \UseAMStwoboldmath
      \makeatletter
      \new@mathgroup\upmath@group
      \define@mathgroup\mv@normal\upmath@group{eur}{m}{n}
      \define@mathgroup\mv@bold\upmath@group{eur}{b}{n}
      \edef\UPM{\hexnumber\upmath@group}
      \new@mathgroup\amsa@group
      \define@mathgroup\mv@normal\amsa@group{msa}{m}{n}
      \define@mathgroup\mv@bold\amsa@group{msa}{m}{n}
      \edef\AMSa{\hexnumber\amsa@group}
      \makeatother
      \mathchardef\upi="0\UPM19
      \mathchardef\umu="0\UPM16
      \mathchardef\upartial="0\UPM40
      \mathchardef\leqslant="3\AMSa36
      \mathchardef\geqslant="3\AMSa3E

      \let\leq=\leqslant \let\le=\leqslant
       \let\ge=\geqslant
    \fi
  \fi
\fi 

\ifnfsstwo
  \DeclareMathAlphabet{\mathbfit}{OT1}{cmr}{bx}{it}
  \SetMathAlphabet\mathbfit{bold}{OT1}{cmr}{bx}{it}
  \DeclareMathAlphabet{\mathbfss}{OT1}{cmss}{bx}{n}
  \SetMathAlphabet\mathbfss{bold}{OT1}{cmss}{bx}{n}
  \ifAMStwofonts
    \ifCUPmtlplainloaded \else
      \DeclareSymbolFont{UPM}{U}{eur}{m}{n}
      \SetSymbolFont{UPM}{bold}{U}{eur}{b}{n}
      \DeclareSymbolFont{AMSa}{U}{msa}{m}{n}
      \DeclareMathSymbol{\upi}{0}{UPM}{"19}
      \DeclareMathSymbol{\umu}{0}{UPM}{"16}
      \DeclareMathSymbol{\upartial}{0}{UPM}{"40}
      \DeclareMathSymbol{\leqslant}{3}{AMSa}{"36}
      \DeclareMathSymbol{\geqslant}{3}{AMSa}{"3E}

      \let\leq=\leqslant \let\le=\leqslant
       \let\ge=\geqslant
    \fi
  \fi
\fi 

\ifCUPmtlplainloaded \else
  \ifAMStwofonts \else 
    \def\upi{\pi}
    \def\umu{\mu}
    \def\upartial{\partial}
  \fi
\fi

\title[LOSVDs of low-luminosity elliptical galaxies]
{Line-of-sight velocity distributions of low-luminosity elliptical
galaxies}

\author[Halliday et~al.] {C. Halliday$^{1,4}$, Roger L. Davies$^1$,
  Harald Kuntschner$^1$, M. Birkinshaw$^2$, \cr Ralf Bender$^3$, R.P.
  Saglia$^3$ and Glenn Baggley$^1$\\
  $^1$ Department of Physics, University of Durham, Science Labs, South
  Road, Durham DH1 3LE, UK.\\
  $^2$ H.H. Wills Physics Laboratory, University of Bristol, Tyndall
  Avenue, Bristol BS8 1TL, UK.\\ 
  $^3$ Universit\"ats-Sternwarte, Scheinerstr. 1, D-81679 M\"unchen,
  Germany.\\ 
  $^4$ Current address: Astrophysics Research Institute, Liverpool John
  Moores University, Twelve Quays House, Egerton
  Wharf,\\ 
  Birkenhead CH41 1LD, UK. email: ch@astro.livjm.ac.uk\\
  }
 
\date{Accepted --- 2001. Received --- ; in original form --- }

\pubyear{2001}

\begin{document}
  
\maketitle
 
\begin{abstract}
  The shape of the line-of-sight velocity distribution (LOSVD) is
  measured for a sample of 14 elliptical galaxies, predominantly
  low-luminosity ellipticals. The sample is dominated by galaxies in
  the Virgo cluster but also contains ellipticals in nearby groups and
  low density environments. The parameterization of the LOSVD due to
  Gerhard and van der Marel \& Franx is adopted, which measures the
  asymmetrical and symmetrical deviations of the LOSVD from a Gaussian
  by the amplitudes $h_{3}$ and $h_{4}$ of the Gauss-Hermite series.
  Rotation, velocity dispersion, $h_{3}$ and $h_{4}$ are determined as
  a function of radius for both major and minor axes. Non-Gaussian
  LOSVDs are found for all galaxies along the major axes. Deviations
  from a Gaussian LOSVD along the minor axis are of much lower
  amplitude if present at all. Central decreases in velocity
  dispersion are found for three galaxies. Two galaxies have
  kinematically-decoupled cores: NGC\,4458 and the well-known case of
  NGC\,3608.
\end{abstract}
 

\section{Introduction}
The shape and strength of absorption features provide a direct measure
of the kinematics of integrated stellar populations in early-type
galaxies \cite{sar77,ton79}. Large format CCD arrays of small pixels
enable two-dimensional studies of galaxy kinematics at high spectral
and spatial resolution to be made using long-slit spectroscopy. This
has led to the study of the line-of-sight velocity distribution
(LOSVD) at different positions within a galaxy, allowing the
kinematical structure to be tightly constrained (e.g., Franx \&
Illingworth 1988, hereafter FI88; Bender 1990, hereafter B90; Rix \&
White 1992, hereafter RW92; van der Marel \& Franx 1993, hereafter
vdMF93; Bender, Saglia \& Gerhard 1994, hereafter BSG94, van der Marel
et al. 1994, hereafter vdM94a, Fisher et~al.  1995, hereafter F95,
Statler \& Smecker-Hane 1999, hereafter SSH99, Mehlert et~al. 2000).

The orbital structure of a galaxy is a fossil record of its
evolutionary history. However, there is a degeneracy between changes in
mass-to-light ratio and velocity dispersion anisotropy that limits the
analysis of rotation and dispersion measurements alone (Binney \&
Tremaine 1987). To recover the mass density distribution of a given
galaxy requires at least an accurate determination of the velocity
dispersion anisotropy through the LOSVD. A full determination of the
galaxy dynamics would require the measurement of surface brightness
profiles and is beyond the scope of this work but an accurate
measurement of the LOSVD is an important step in achieving this (e.g.,
van der Marel et~al. 1994, hereafter vdM94b; Saglia et al. 2000).

\begin{figure}
{\epsfxsize=8.3truecm \epsfysize=8.3truecm 
\epsfbox[10 150 570 660]{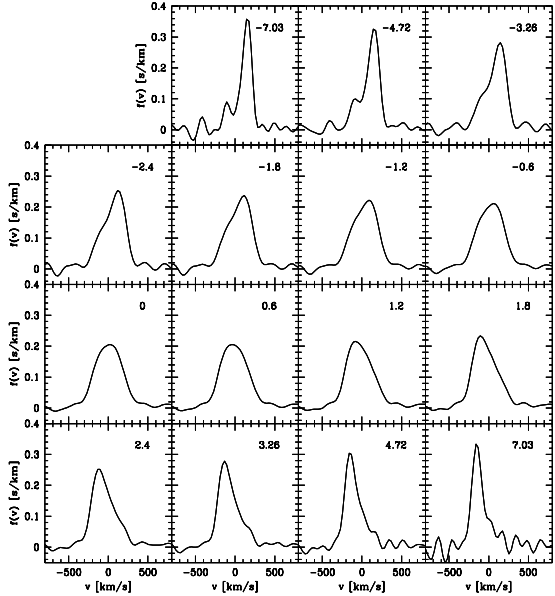}}
\caption
{\small \label{5582mj_BF} Recovered galaxy broadening functions for
the major axis spectrum of NGC\,5582. Radii of measurement are
indicated in arcsec in each panel. The shape of the broadening
function is clearly non-Gaussian for most of the radial bins. It is
also evident how the width of the LOSVD increases towards smaller
radii, in agreement with the larger measured values of the velocity
dispersion $\sigma$.}
\end{figure}

To a first approximation, elliptical galaxies are believed to be
relaxed dynamical systems and the distribution of stellar velocities
has been assumed to be Maxwellian producing approximately Gaussian
profiles. However, within a number of ellipticals non-Gaussian LOSVDs
have been measured. For example, see Figure~\ref{5582mj_BF} for LOSVDs
measured in NGC\,5582 at various radii. The non-Gaussian LOSVDs led to
the detection of multiple component structures in ellipticals, such as
kinematically-decoupled cores (FI88; Jedrzejewski \& Schechter 1988,
hereafter JS88), and central disks (RW92; van den Bosch et al. 1998).
It has been demonstrated that under the assumption of Gaussian LOSVDs,
the measurement of both rotation velocity and velocity dispersion can
be incorrect by 10\% or more (vdMF93).

To improve on the earlier measurements, more general parameterizations
of the LOSVD have been tested and have been found to provide a superior
description of the galaxy kinematical structure (RW92; vdMF93; Kuijken
\& Merrifield 1993; Zhao \& Prada 1996). Here galaxy kinematics are
measured using the Fourier Correlation Quotient (FCQ) method of B90.
This method is considerably less sensitive to template-mismatching than
traditional Fourier methods and allows the measurement of the full
LOSVD.  The LOSVD is then parameterized as a Gauss-Hermite series
expansion (vdMF93) including the higher order terms $h_{3}$ and
$h_{4}$.

The galaxy sample studied here consists mostly of low-luminosity
ellipticals (hereafter LLEs). Evidence has grown steadily over the
past 15 years that LLEs may be very different from the more luminous
giant ellipticals, both in their kinematical behaviour (Davies
et~al. 1983, hereafter DEFIS83; Bender \& Nieto 1990; Rix et~al.
1999) and the properties of their stellar populations (e.g. Faber
et~al. 1992). As isotropic rotators (DEFIS83), LLEs are more easily
modelled than the giant ellipticals which are believed to be supported
by velocity dispersion anisotropy. Furthermore, the lower velocity
dispersion broadening of LLEs enables the higher-orders of the LOSVD
to be measured more accurately than for the more luminous ellipticals,
provided that data of sufficiently high spectral resolution and
signal-to-noise ratio (hereafter S/N) can be obtained.

The main objective of the measurements described here is to constrain
the kinematical structure of our LLE sample as a function of radius. In
Section~\ref{obs}, observations and reduction steps are described. The
adopted LOSVD parameterization is presented and discussed in
Section~\ref{losvdpar}. A description of the measurement of galaxy
kinematics using the FCQ method and measurements of rotation, $\sigma$,
$h_{3}$ and $h_{4}$ are given in Section~\ref{kinmeas} and Appendix~A.
A short analysis of characteristic global kinematic parameters for the
major axis of our sample is presented in Section~\ref{sec:charpar} and
a discussion of our results is given in Section~\ref{sec:discussion}.
Comparisons between measurements obtained here and the measurements of
previous authors are presented in Appendix~B.

\section{Observations and Basic Reduction}\label{obs}
\subsection{Galaxy Sample}\label{obsred}
A complete sample of 17 low-luminosity elliptical galaxies was selected
from the Virgo catalog of Bingelli et~al. (1985) covering the absolute
magnitude range $-20 \leq M_{B} \leq -17$.

Nine galaxies of this original sample were observed. Other
low-luminosity ellipticals (NGC\,2778, NGC\,3377, NGC\,3605, NGC\,3608,
NGC\,5582) were observed when Virgo was not available due to cloudy or
windy conditions. The well studied `bona fide' elliptical NGC\,3379 was
observed to enable consistency checks to be made with the measurements
of other authors. Our observed sample is given in Table~\ref{obssmp}
which lists the morphological type, apparent magnitude B$_T$, the
effective radius in arcsec (R$_e$), mean ellipticity, heliocentric
redshift, cz$_{helio}$ and environment. Values of absolute magnitude,
as listed also in Table~\ref{obssmp}, were calculated using the
distances from the surface-brightness-fluctuation survey of Tonry
et~al. (2001).  Specifically we use a mean distance of 17.0~Mpc to the
Virgo cluster, 11.1~Mpc to the Leo~I group and 21.8~Mpc to the Leo~II
group. Two galaxies, NGC\,2778 and NGC\,5582, are in low density
environments and we use the individual SBF distances of 16.9~Mpc and
28.4~Mpc, respectively.

\begin{table*}
\caption{Observed Galaxy Sample}
\label{obssmp}
\begin{tabular}{clcrrrcc} 
\hline
    Name   &Type& B$_T$ & R$_e$ & $\overline{\epsilon}$  & cz$_{helio}$ & M$_B$  & environment \\ 
     (1)   & (2)&  (3)  &  (4)  &  (5)        &  (6)         &   (7)  &   (8) \\ \hline
 NGC\,2778  & E  & 13.35 & 15.74 & 0.22       & 2019         & -18.5  & - \\
 NGC\,3377  & E5+& 11.24 & 34.45 & 0.48       &  692         & -19.0  & Leo I group\\
 NGC\,3379  & E0 & 10.24 & 35.25 & 0.09       &  889         & -20.0  & Leo I group\\
 NGC\,3605  & E4+& 13.13 & 21.24 & 0.34       &  649         & -18.6  & Leo II group \\
 NGC\,3608  & E2 & 11.70 & 33.66 & 0.21       & 1205         & -20.0  & Leo II group \\
 NGC\,4239  & E  & 13.70 & 15.39 & 0.46       &  921         & -17.5  & Virgo \\
 NGC\,4339  & E0 & 12.26 & 32.15 & 0.08       & 1281         & -18.9  & Virgo \\
 NGC\,4387  & E  & 13.01 & 15.74 & 0.35       &  561         & -18.1  & Virgo \\
 NGC\,4458  & E0+& 12.93 & 26.13 & 0.13       &  668         & -18.2  & Virgo \\
 NGC\,4464  & S? & 13.46 &  7.54 & 0.24       & 1255         & -17.7  & Virgo \\
 NGC\,4467  & E2 & 14.77 & 10.64 & --         & 1474         & -16.4  & Virgo \\
 NGC\,4468  & S0?& 13.58 & 27.36 & 0.29       &  895         & -17.6  & Virgo \\
 NGC\,4478  & E2 & 12.36 & 13.40 & 0.17       & 1382         & -18.8  & Virgo \\
 NGC\,4551  & E  & 12.97 & 13.10 & 0.23       & 1189         & -18.2  & Virgo \\
 NGC\,4564  & E6 & 12.05 & 19.82 & 0.58       & 1119         & -19.1  & Virgo \\
 NGC\,5582  & E  & 12.48 & 32.89 & 0.32       & 1332         & -19.8  & - \\ \hline
\end{tabular}
\medskip
\begin{minipage}{15.4cm}
  {\em Notes:}\/ Values of B$_T$ in magnitudes, effective radius R$_e$
  in arcsec, and heliocentric redshift cz$_{helio}$ in km s$^{-1}$
  were taken from the Third Reference Catalogue (de Vaucouleurs
  et~al. 1991, hereafter RC3). The Hubble type, also taken from the
  RC3, is given in column 2: E-elliptical, S-spiral, S0-lenticular,
  +-intermediate (i.e. between elliptical and S0), ?-doubtful
  classification. Column 5 provides the mean ellipticity used in
  Section~\protect\ref{sec:charpar}: these values were taken from
  Peletier et~al. (1990), Bender et~al. (1989) and Lauer et
  al. (1995); for NGC's 4239, 4339, 4458, 4468, 5582 values were
  obtained from the Digital Sky Survey. \protect\footnote{The
  Digitized Sky Surveys were produced at the Space Telescope Science
  Institute under U.S. Government grant NAG W-2166.} To calculate the
  absolute magnitude M$_B$ (column 7) we used the distances based on
  the surface-brightness-fluctuation survey of Tonry et~al. (2001).
  Galaxies from Bingelli et~al. (1985) that were contained in our
  original sample, but could not be observed are NGC\,4168, NGC\,4261,
  NGC\,4360, NGC\,4486A, NGC\,4515, IC 3653, NGC 4623, NGC\,4660.
  Column~8 indicates whether each galaxy is a member of the Virgo
  cluster, the Leo~I group, the Leo~II group or resides in a low
  density environment (-).
\end{minipage}
\end{table*}

\subsection{Description of Observations}\label{descobs}
Long-slit spectroscopic data were obtained using the Blue Channel
Spectrograph at the Multiple Mirror Telescope, Arizona, U.S.A., during
5-7 March and 15-17 May 1994, and 22-26 February 1995. The seeing
varied from 0.5 to 2.0 arcsec. The Loral CCD detector of the Blue
Channel Spectrograph was replaced between the observing runs of May
1994 and February 1995 with a chip of improved cosmetics. A summary of
the instrumental set-up is given in Table \ref{chip}.

\begin{table}
\caption{The instrumental set-up} 
\label{chip}
\begin{tabular}{ll}
\hline
 Telescope         & MMT                       \\
 Instrument        & MMT spectrograph, blue channel   \\
 Grating           & 1200 g/mm                 \\
 Detector          & Loral (3k$\times$1k)             \\
 Pixel Size        & 15$\mu$m x 15$\mu$m       \\
 Readout Noise     & $\sim 7-8$~e$^{-}$          \\
 Gain              & $\sim 1.5$~e$^{-}$/ADU    \\
 Slit width        & 1\arcsec                      \\
 Spatial Scale     & 0\farcs6/pixel $^{a}$     \\
 Instr. resolution & $\sim$1.5~\AA\/ (FWHM)       \\ 
 Dispersion        & 0.5~\AA/pixel             \\ 
 Wavelength range  & 4555 - 6045~\AA           \\ \hline
\multicolumn{2}{l}{$^{a}$ binned $2 \times$ minimum bin size in spatial direction}\\
\end{tabular}
\end{table}

A wavelength range of $\sim$4555-6045~\AA\/ was chosen to be centred on
the Mg\,$b$, Fe5270 \& Fe5335 Lick/IDS line-strength indices
\cite{tra98}. To accurately determine values of $\sigma$ of as low as
60 km s$^{-1}$, a high spectral resolution of $\sim$1.5~\AA\/ FWHM was
required; for this purpose a grating of 1200 g/mm and a slit width of
1~arcsec were used. The resolution element in the spatial direction was
rebinned by 2 pixels to produce a spatial scale of 0\farcs6
pixel$^{-1}$. The slit length was 180 arcsecs.

High S/N long-slit spectra were obtained for the major and minor axes
of each galaxy of our sample. To determine kinematics 3-8 stellar
template stars, predominantly G and K giants, were observed during each
night. A log of all galaxy observations is given in Table~\ref{gallog}:
this includes the date of observation, position angle (hereafter PA) of
observation and the exposure time in seconds. Data for the galaxies
NGC\,4467 and NGC\,4239 were of poor S/N and results for these objects
will not be presented here.

\begin{table}
  \caption{Observing Log}
  \label{gallog}
  \begin{tabular}{ccrr} 
    \hline 
    Date     & Name & PA (deg.) & Exp. (s) \\
      (1)    &   (2)     &  (3)    & (4)   \\ \hline
    05.03.94 & NGC\,3605  & 290 min & 3600  \\
    05.03.94 & NGC\,3605  & 200 maj & 3600  \\
    05.03.94 & NGC\,4387  & 322 maj & 3600  \\
    05.03.94 & NGC\,4387  &  52 min & 3600  \\
    05.03.94 & NGC\,4564  & 135 min & 3600  \\
    05.03.94 & NGC\,4468  & 155 min & 3600  \\
    06.03.94 & NGC\,3608  & 351 min & 3600  \\
    06.03.94 & NGC\,2778  & 225 maj & 3600  \\
    06.03.94 & NGC\,2778  & 135 min & 3600  \\
    06.03.94 & NGC\,5582  & 205 maj & 3600  \\
    15.05.94 & NGC\,4564  &  45 maj & 3600  \\
    15.05.94 & NGC\,4468  &  65 maj & 3600  \\
    15.05.94 & NGC\,4464  &  95 min & 3600  \\
    15.05.94 & NGC\,4464  &   5 maj & 3600  \\
    15.05.94 & NGC\,5582  & 295 min & 4200  \\
    16.05.94 & NGC\,3608  &  81 maj & 3600  \\
    22.02.95 & NGC\,4467  &  41 maj & 3600  \\
    22.02.95 & NGC\,4467  & 131 min & 3600  \\
    23.02.95 & NGC\,3379  &  71 maj &  600  \\
    23.02.95 & NGC\,3379  & 161 min &  600  \\
    23.02.95 & NGC\,3377  &  42 maj & 1800  \\
    23.02.95 & NGC\,3377  & 312 maj & 1800  \\
    23.02.95 & NGC\,4551  &  70 maj & 3600  \\
    23.02.95 & NGC\,4551  & 340 min & 3600  \\
    23.02.95 & NGC\,4458  &   5 maj & 3600  \\
    23.02.95 & NGC\,4458  &  95 min & 3600  \\
    23.02.95 & NGC\,4478  & 145 maj & 3600  \\
    24.02.95 & NGC\,4478  &  55 min & 3600  \\ 
    24.02.95 & NGC\,4339  &  20 maj & 2700  \\ 
    24.02.95 & NGC\,4339  & 110 min & 2700  \\ 
    24.02.95 & NGC\,4239  & 115 maj & 5400  \\ \hline
\end{tabular} 
\medskip
\begin{minipage}{7cm} 
  {\em Notes:}\/ The date of each observation and galaxy name is given
  in columns 1 \& 2, respectively. Column 3 gives the position angle of
  the observation with sense North through East; positions
  corresponding to the major (maj) and minor (min) axes are indicated.
  Column 4 lists the exposure time in seconds.
\end{minipage}
\end{table}

Two different types of stellar observation were made: `trailed' and
`rocked'. `Trailed' spectra were obtained by trailing the stellar
image along the length of the slit while the telescope was slightly
de-focussed. A `rocked' spectrum was created by moving the stellar
image {\it across} the slit at regular intervals (10-20 arcsecs) along
the slit length. This helped to ensure that the illumination of the
slit and hence the instrumental broadening of observations, was
identical for both star and galaxy spectra, and for this reason
`rocked' spectra were preferred for the kinematical analysis.

Exposures of both an incandescent lamp and the twilight night sky were
obtained as flat-field standards. Standard He-Ar and Ne comparison
lamp sources were observed simultaneously to enable wavelength
calibration and geometrical rectification to be completed for all
stellar and galaxy observations. For galaxy observations, comparison
exposures were taken both before and after the exposure to monitor any
shift in the instrumental set-up. For a given star, a single
comparison exposure of $\sim$2-4 minutes was obtained for {\it both}
the `rocked' and `trailed' exposures where these were taken
consecutively.

\subsection{Basic Reduction}\label{red}
A standard data reduction was performed principally using IRAF
\footnote{Image Reduction and Analysis Facility of the National
Optical Astronomy Observatories, Tuscon, Arizona, U.S.A..}. To
complete the measurement of galaxy kinematics additional steps were
performed using IRAF and MIDAS \footnote{Munich Image Data Analysis
System of the European Southern Observatory.}.

Reduction steps, performed for each night of observation, are
summarized; for a more detailed description see Halliday (1998). All
CCD observations were bias-subtracted, flat-fielded and corrected for
the effects of both cosmic ray hits and bad pixel columns. A wavelength
calibration was performed for each star and galaxy frame to a typical
accuracy of 0.05~\AA\/ ($\sim$ 3 km s$^{-1}$ at $\sim$5200~\AA). A
series of distortion maps were created to geometrically rectify each
frame to a typical accuracy of 0.2 pixels RMS ($\sim$6 km s$^{-1}$ at
$\sim$5200~\AA). Background sky signal was subtracted for each galaxy
frame by linearly-interpolating between areas of sky at opposite sides
of the galaxy spectrum. A correction for chromatic focus variations
within a galaxy frame was made where required. This involved measuring
the central width of the galaxy profile as a function of wavelength. A
procedure was developed to smooth the galaxy profile width to the same
value for all wavelengths.  This was similar to a procedure adopted by
Mehlert et~al. (1998) to correct for chromatic focus variations
detected within their data. The dispersion axis of each galaxy frame
was finally rebinned on to a natural logarithmic scale.

A one-dimensional stellar spectrum was created for each `rocked' star
frame by extracting individual `strips' of spectra within each
frame. These were combined to produce a single one dimensional
spectrum and deredshifted to laboratory wavelengths.

\section{Parameterization of the Line-of-Sight Velocity Distribution}\label{losvdpar}
The line-of-sight velocity profile was decomposed using the
parameterization proposed by Gerhard (1993) and van der Marel and Franx
(1993):

\begin{equation}\label{losvd}
 f(y) =I_0 e^{-\frac{y^2}{2}}(1+h_3 H_3 (y)+h_4 H_4 (y))
\end{equation}

\noindent where $ y={(\nu_{fit}-\nu)\over\sigma_{fit}}$, $\nu_{fit}$
is the measured velocity; $\nu$ is the mean of all measured velocities
for a given spectrum (i.e., the mean radial velocity); $\sigma_{fit}$
is the measured velocity dispersion; $H_3 (y)$ and $H_4(y)$ are
antisymmetric and symmetric standard Gauss-Hermite polynomials of 3rd
and 4th order and $h_3$ and $h_4$ are their coefficients, respectively;
and $I_0$ is a normalization constant.

The coefficient $h_3$ quantifies the LOSVD asymmetry. Measured
asymmetries for elliptical galaxies arise from a non-Gaussian velocity
distribution along the line-of-sight, e.g., the superposition of a
slowly rotating bulge and a more rapidly rotating disk component.
Positive measurements of $h_3$ correspond to a distribution skewed
towards velocities {\it lower} than the measured systemic velocity of
the galaxy. Conversely negative $h_3$ measurements correspond to
distributions skewed towards velocities greater than the systemic
velocity.

The $h_4$ term quantifies the symmetric deviations of the LOSVD from a
Gaussian. For $h_4 > 0$ the corresponding distribution is more peaked
than a Gaussian at small velocities with more extended high velocity
tails. Conversely, distributions less peaked than a Gaussian will have
$h_4 < 0$.

\section{Galaxy Kinematics: Measurement and Results}\label{kinmeas}
Measurements of rotation, $\sigma$, $h_3$ and $h_4$ were obtained
using the FCQ method of B90. This procedure involved extraction of an
appropriate wavelength range, binning the spectra as a function of
radius to a particular value of S/N, continuum level removal and
application of a Wiener filter.

Optimal values of parameters for each of the measurement steps were
established using Monte-Carlo simulations. Simulated `galaxy' spectra
for particular measurements of $\sigma$, $h_3$ and $h_4$ were created
from one-dimensional stellar spectra. To model the effect of noise in
simulations for a given S/N, different Gaussian noise realisations were
added to 30 copies of each simulated galaxy spectrum. The mean value
and spread of kinematical measurements obtained for the different
spectra were used to determine the optimal value for each parameter
(for details see Halliday 1998).

From our overall wavelength range, we extracted a range centred on the
Mg\,$b$ spectral feature and excluding H$\beta$. Galaxy spectra were
rebinned in the direction of the spectroscopic slit to a minimum
signal-to-noise ratio (S/N) of 60 per \AA\/ and separately for S/N = 30
per \AA. The shapes of continuua were removed for the stellar and
galaxy spectra by fitting and dividing by a third order polynomial. The
edges of all spectra were tapered using a cosine bell function.
Filtering of high frequency noise was performed using an appropriate
Wiener filter. Error estimation involved the calibration of the
peak-to-noise ratio of the power-spectrum of the galaxy spectra with
the value of the signal-to-noise ratio. Simulations of galaxy spectra
at different radii were performed to derive their signal-to-noise ratio
and therefore the error values.

Measurements were obtained for 3-8 different stellar templates. Only
templates observed on the same night of observation were used for a
given galaxy; this is important since a central assumption of the FCQ
method is that the instrumental broadening of the galaxy and stellar
spectra can be assumed to be similar. Although the FCQ method reduces
the effect of template-mismatching, measurements of the higher order
terms $h_{3}$ and $h_{4}$ are still sensitive to spectral differences
between template star and galaxy. For an elliptical galaxy of
axisymmetric kinematical structure, where the spectroscopic slit has
been carefully centred on the galaxy centre, the $h_{3}$ profile is
expected to have reflection symmetry about the galaxy centre.
Template-mismatching produces a constant offset in $h_{3}$. The final
template was choosen to minimize the offset in $h_{3}$. Where possible,
identical templates were used to derive measurements for the major and
minor axis spectrum of each galaxy. Where major and minor axis
observations were obtained on different nights and identical templates
could not be used, templates of most closely matching stellar type were
chosen.

In Appendix \ref{kinmeasplot} we present measurements of rotation,
$\sigma$, \hth\/ and \hf\/ for the major and minor axes of our sample
galaxies. A range of different kinematical behaviour is evident.
Significant LOSVD asymmetries are measured for 13 out of 14 galaxies
along the major axis. Central dips in $\sigma$ are measured for the
major and minor axis of three galaxies. Two galaxies are found to have
kinematically-decoupled cores.

\section{Characteristic Kinematical Parameters}
\label{sec:charpar}
In this section characteristic, global kinematic parameters are
derived and analysed for the major axes of each galaxy. Six parameters
were determined for each individual galaxy: $v_{max}$ the maximum
measured rotation velocity; $\sigma_{0}$, $\sigma_{max}$ and
$\overline\sigma$, the central, maximum and mean velocity dispersion,
respectively. For $\overline\sigma$, values were taken to be the mean
of all measurements within $\frac{1}{2} $R$_{e}$. Furthermore
${\overline{|h_{3}|}}$, the mean of the absolute values of $h_{3}$,
and ${\overline{h_{4}}}$, the mean value of $h_{4}$ were
determined. For ${\overline{|h_{3}|}}$ and ${\overline{h_{4}}}$, the
mean of all measurements within $\frac{1}{2} $R$_{e}$, excluding the
innermost measurements ($|r| \la 2$\farcs$6$), was
taken. Uncertainties were calculated by propagating the errors
determined for the original measurements of rotation, $\sigma$,
$h_{3}$ and $h_{4}$. A summary of all values is given in
Table~\ref{tab:para}.

\begin{table*}
\caption{Global parameters}
\label{tab:para}
\begin{tabular}{crrrrrrrrcccc} 
\hline
Name & $v_{max}$ & d$v_{max}$ & $\sigma_{max}$ & d$\sigma_{max}$ &
$\sigma_{0}$ & d$\sigma_{0}$ & $\overline{\sigma}$ & 
d$\overline{\sigma}$ & $\overline{|h_3|}$ & d$\overline{|h_3|}$ & $\overline{h_4}$ & d$\overline{h_4}$\\ 
     (1) & (2) & (3) & (4) & (5) & (6) & (7) & (8) & (9) & (10) & (11) & (12) & (13) \\ \hline

NGC\,2778& 109.4 & 9.3 & 169.2 & 1.0 & 168.7 & 0.9 &  142.1   & 0.6   & 0.078 & 0.008 & 0.028 & 0.009\\
NGC\,3377& 100.1 & 0.9 & 156.0 & 1.8 & 156.0 & 1.8 &  101.6   & 0.5   & 0.124 & 0.006 &-0.035 & 0.009\\
NGC\,3379&  50.0 & 5.7 & 224.5 & 2.8 & 221.1 & 2.9 &  209.5   & 1.3   & 0.025 & 0.009 &-0.001 & 0.010\\
NGC\,3605&  55.9 & 8.5 &  91.9 & 0.6 &  90.8 & 0.5 &   87.6   & 0.2   & 0.042 & 0.005 & 0.002 & 0.005\\
NGC\,3608&  45.4 & 5.2 & 201.6 & 0.8 & 201.6 & 0.8 &  179.3   & 0.4   & 0.034 & 0.003 & 0.016 & 0.003\\
NGC\,4339&  56.3 & 5.4 & 115.2 & 1.0 & 115.2 & 1.0 &  100.4   & 0.5   & 0.072 & 0.007 & 0.081 & 0.010\\
NGC\,4387&  59.8 & 7.4 & 101.4 & 0.9 &  97.9 & 0.6 &   97.1   & 0.3   & 0.013 & 0.003 & 0.009 & 0.004\\
NGC\,4458&  26.3 & 1.2 & 119.0 & 2.1 & 119.0 & 2.1 &   97.8   & 0.7   & 0.059 & 0.013 &-0.035 & 0.018\\
NGC\,4464&  78.0 & 1.2 & 141.2 & 0.6 & 141.2 & 0.6 &  120.9   & 0.4   & 0.150 & 0.007 & 0.051 & 0.008\\
NGC\,4468&  24.2 & 2.9 &  36.2 & 1.6 &  34.6 & 1.2 &   33.2   & 0.8   & 0.029 & 0.043 &-0.060 & 0.027\\
NGC\,4478&  58.4 & 3.1 & 146.0 & 3.7 & 126.3 & 2.0 &  131.7   & 0.8   & 0.054 & 0.007 &-0.014 & 0.008\\
NGC\,4551&  46.6 & 2.0 & 103.5 & 1.2 & 103.5 & 1.2 &   95.2   & 0.4   & 0.053 & 0.005 &-0.034 & 0.008\\
NGC\,4564& 153.2 & 1.6 & 177.4 & 0.4 & 177.4 & 0.4 &  144.8   & 0.2   & 0.027 & 0.002 &-0.016 & 0.002\\
NGC\,5582& 165.2 &10.0 & 164.6 & 0.7 & 164.6 & 0.7 &  125.8   & 0.5   & 0.169 & 0.007 & 0.104 & 0.010\\\hline
\end{tabular}
\medskip
\begin{minipage}{15cm}
  Global parameters, as plotted in figures \protect{\ref{vsig}} and
  \protect{\ref{kin_typ}}, are given for the major axis spectrum of
  each galaxy. Columns 2 \& 3 give the maximum measured rotation
  velocity and its error. The maximum, central and mean velocity
  dispersion and their corresponding errors are given in columns 4 \&
  5, 6 \& 7 and 8 \& 9, respectively. Columns 10 \& 11 give the mean
  absolute value of $h_3$ and its error. The mean value of $h_4$ and
  its error are given in columns 12 \& 13. All measurements of velocity
  and $\sigma$ are given in \kms.
\end{minipage}
\end{table*}

\begin{figure*}
\begin{center}
\begin{minipage}{6in}
    \epsfig{file=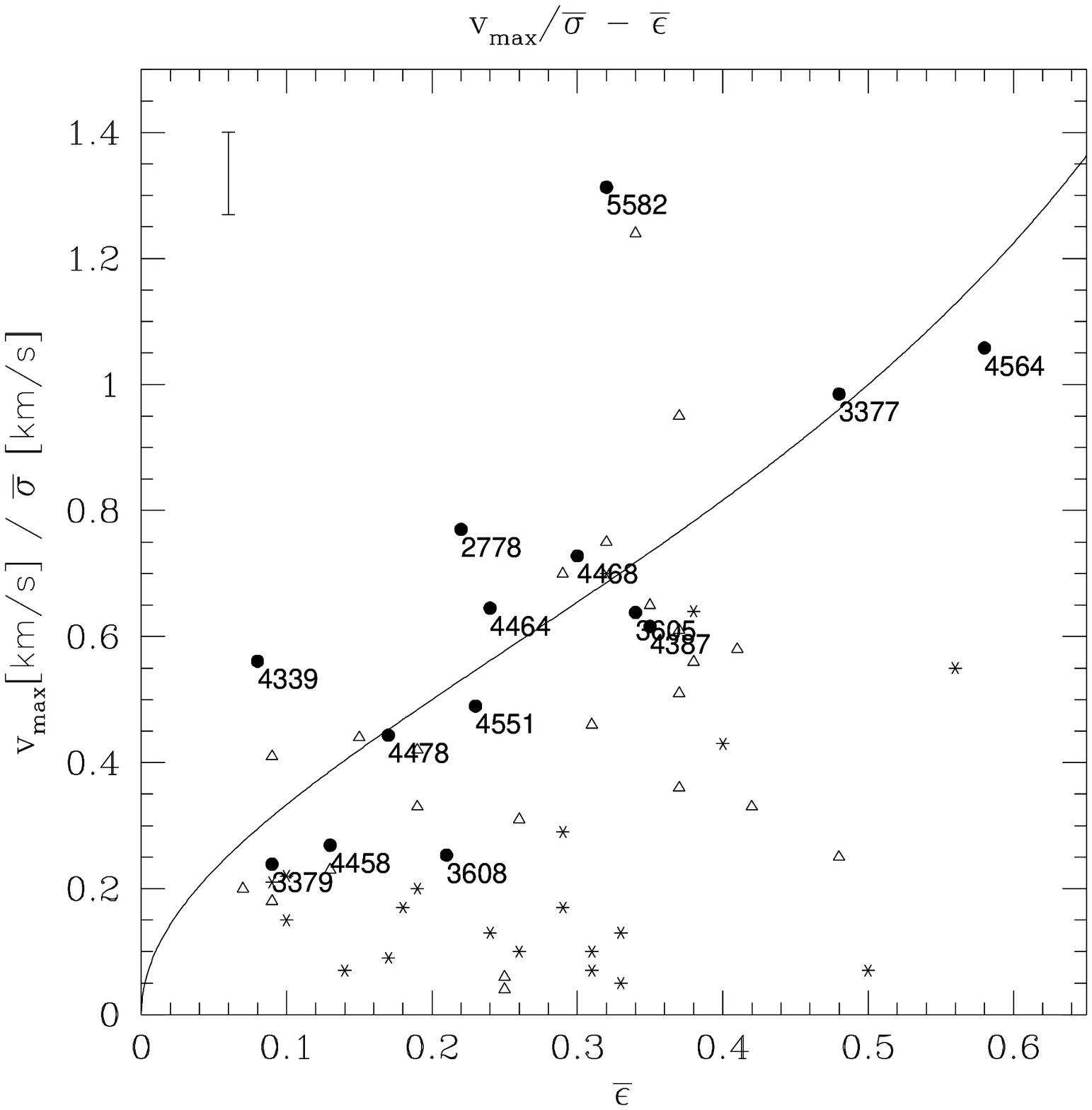,width=15.0cm}
\end{minipage} 
\caption{\small \label{vsig} Comparison of measurements of
  $v_{max}$~/~$\overline{\sigma}$ against the mean ellipticity
  $\overline\epsilon$, with predictions for oblate isotropic models
  (Binney 1978, solid line). Measurements for galaxies studied here
  are indicated by filled circles. A typical error bar for
  $v_{max}$~/~$\overline{\sigma}$ is given in the upper left-hand
  corner. Additional measurements are from DEFIS83: measurements for
  giant ellipticals are given by asterisk symbols, and low-luminosity
  ellipticals are represented by open triangles.}
\end{center}
\end{figure*}

In Figure~\ref{vsig} the ratio of $v_{max}$ to the mean velocity
dispersion $\overline{\sigma}$ is plotted against the mean ellipticity
$\overline\epsilon$. Predictions for an oblate isotropic rotator model
from Binney (1978) are shown as a solid line. These predictions were
obtained using the tensor virial theorem, assuming constant
ellipticity. Measurements obtained here for the low luminosity
ellipticals are found close to the predictions for isotropic rotators
with the exception of NGC\,5582. The latter galaxy shows unusually
strong rotation for its mean ellipticity. Our data is in good agreement
with the earlier findings of DEFIS83 (their LLEs are shown as open
triangles in Figure~\ref{vsig}).

\begin{figure*}
\begin{center}
\begin{minipage}{6in}

\epsfig{file=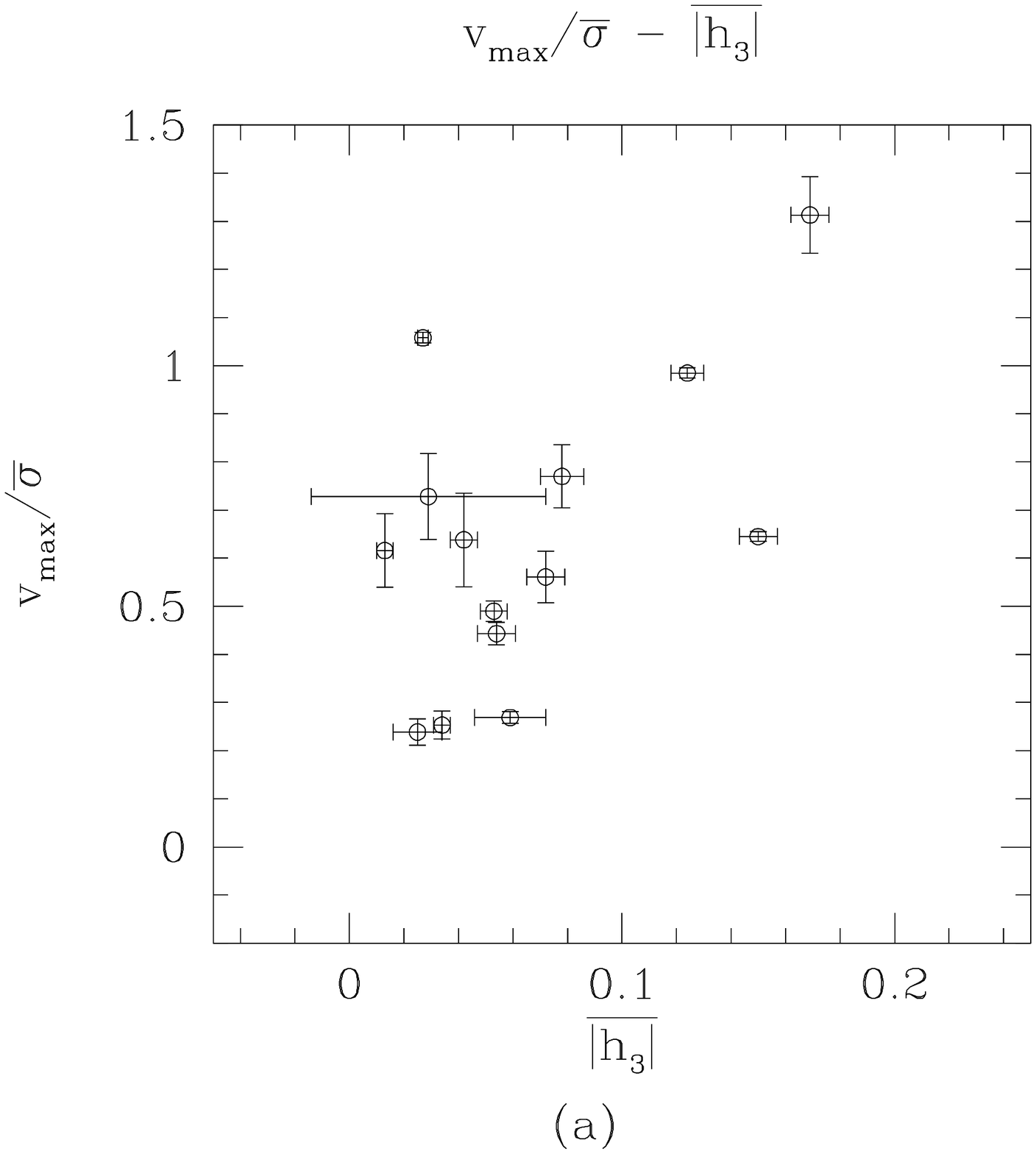,width=7.5cm}
\epsfig{file=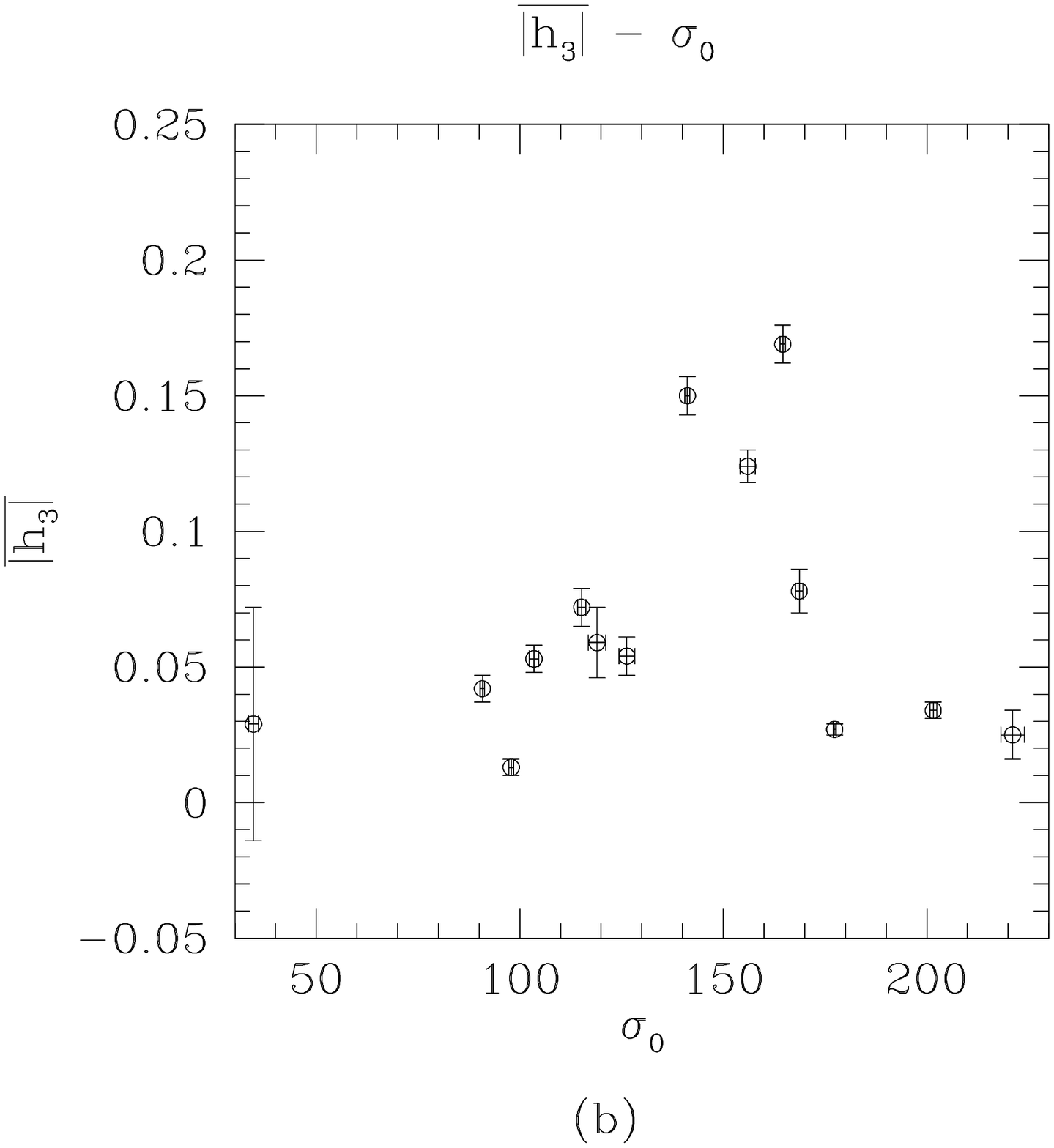,width=7.5cm}
\end{minipage} 
\caption{\small \label{kin_typ} Relationships between global 
  parameters for the major axis of each galaxy. (a) $v_{max} /
  \overline{\sigma}$ {\em vs}\/ $|\overline{h_3}|$; (b)
  $|\overline{h_3}|$ {\em vs}\/ $\sigma_0$. Values of $v_{max}$,
  \protect$\overline\sigma$ and $\sigma_{0}$ are given in units of km
  s$^{-1}$.}
\end{center}
\end{figure*}

The deviations from a Gaussian LOSVD as measured by the higher order
terms \hth\/ and \hf, span a large range in values within and between
galaxies. The mean values of $|\overline{h_3}|$ vary from $+0.013$ to
$+0.169$ between galaxies, and $\overline{h_4}$ varies from $-0.060$
to $+0.104$. The largest values of \hth\/ and \hf\/ are detected in
NGC\,5582. We did not find clear correlations between our global
parameters and the higher order terms \hth\/ and \hf. However, two
interesting diagrams concerning \hth\/ are shown in
Figure~\ref{kin_typ}.

In Figure \ref{kin_typ}(a) the ratio of $v_{max}$ to
$\overline{\sigma}$ is plotted against ${\overline{|h_{3}|}}$ for each
galaxy. We note that there is a weak trend such that galaxies with
larger ratios of $v_{max}/\overline{\sigma}$ have on average larger
values of ${\overline{|h_{3}|}}$. This suggests that greater rotational
support may be related to larger LOSVD asymmetries. In Figure
\ref{kin_typ}(b) it is shown that the largest values of
${\overline{|h_{3}|}}$ are measured for galaxies with intermediate
values of $\sigma_{0}$. LOSVD asymmetries may be more common for these
galaxies. A larger sample is needed to confirm this suggestion.

\section{Discussion}
\label{sec:discussion}
We have demonstrated in Section~\ref{sec:charpar} that for our sample
of LLEs the {\em global}\/ kinematic parameters such as $v_{max}$,
$\overline{\sigma}$ and $\overline{\epsilon}$ are consistent with the
predictions for a simple oblate isotropic rotator model.

In general, however, we find a wide range of complex kinematical
behaviour in our sample. There is a wide range of velocity dispersion
gradients with radius. Some galaxies, such as NGC\,3377, show a steep
decline, whereas others, such as NGC\,4551, have a rather shallow
profile.  Central decreases in $\sigma$ are measured for three
galaxies for the major and minor axes (NGC\,4387, NGC\,4478 and
NGC\,4551), and two galaxies (NGC\,3608 and NGC\,4458) are found to
have kinematically-decoupled cores (KDCs). Significant asymmetric
line-of-sight velocity distributions, parameterized by the
Gauss-Hermite coefficient $h_{3}$, are measured for 13 low-luminosity
ellipticals. These asymmetries are found mainly for radii close to the
galaxy centre, and suggest that we are detecting the motion of two or
more components of different kinematical properties in at least some
of the low-luminosity ellipticals.

The LOSVDs along the minor axis are generally consistent with a
Gaussian shape. Only a few galaxies show significant \hth\/ or \hf\/
terms. In the cases where non-Gaussian LOSVDs are detected this can be
explained by isophote twisting such that our slit does not follow the
exact location of the kinematic minor axis. This may also explain the
non-zero velocities measured for some galaxies along the minor axis
(e.g, NGC\,3377 and NGC\,4464). NGC\,4478 is the most spectacular
case, since it exhibits signs of disturbed, perhaps unrelaxed
kinematics. This galaxy is not axisymmetric and any dynamical model
would need to be triaxial. This is also likely to be the case for the
other galaxies which show non-zero velocities along the minor axis.

Some of the kinematic signatures we detect in our sample of LLEs, such
as KDCs and central dips in velocity dispersion, are indicative of a
merger scenario. It is interesting to note that in their recent merger
simulations, Bendo and Barnes (2000, hereafter BB2000) have presented
LOSVDs for the merger remnants of two spiral galaxies, which resemble
elliptical galaxies. In these simulations central dips in $\sigma$
were found for remnants of an {\em equal-mass}\/ merger of two
spirals. Predictions for the profiles of rotation, $h_3$ and $h_4$ are
however different from those observed in our study. In most cases
$h_3$ is predicted to have a similar sign to the rotation velocity
close to the galaxy centre; for our galaxy sample, $h_3$ has an {\em
opposite} sign to rotation. In one case in the study of BB2000, where
$h_3$ and rotation do have opposite signs, the merger remnant has a
counterrotating core: neither of the two galaxies in our sample with a
counterrotating core (i.e. NGC 3608 and NGC 4458) have a central
decrease in $\sigma$. For NGC 4387, rotation rises steadily outside
the galaxy centre in agreement with the simulation results, although
$h_4$ is consistent with zero in contrast to the predicted values of
positive $h_4$. Rotation velocities for the unequal-mass mergers
considered by BB2000 are of far greater amplitude than the velocities
measured in this study.

Our long-slit observations provide a valuable insight into the
dynamics of LLEs and demonstrate clearly that there exists a wide
range of complex, sometimes non-axisymmetric kinematics. With data
available only for the major and minor axis of a galaxy it is
difficult, if not impossible, to describe the full dynamical make-up
of galaxies. The newly commisioned integral field units such as SAURON
(Bacon et al.  2001), however, provide 2-dimensional rotational
velocity and velocity dispersion maps of early-type galaxies and will
advance this area of astronomy.

\section{Acknowledgements}
We thank the CFA Observing Time Committee for their generous
allocation of telescope time. CH thanks Frank van den Bosch for kindly
providing his photometric measurements for NGC\,4478 and NGC\,4564. CH
acknowledges helpful discussions with Reynier Peletier and the support
of a PPARC studentship at the University of Durham and a PPARC PDRA
grant at Liverpool JMU. HK was supported at the University of Durham
by a PPARC rolling grant in Extragalactic Astronomy and Cosmology.

\appendix\label{kinres}
\section{Measurements of Galaxy Kinematics}\label{kinmeasplot}
We present our measurements for v, $\sigma$, $h_3$ and $h_4$ as a
function of radius for all galaxies in order of their NGC number.

The systemic velocity of each galaxy was removed by subtracting the
mean velocity. Small shifts ($<$8~\kms) were applied in order to centre
the velocity profile about the apparent kinematic centre of the galaxy.
The PA of observation is indicated in each case (values of positive
radius correspond to positions in the direction of the PA).
Measurements of rotation, $\sigma$, \hth\/ and \hf\/ obtained for a S/N
binning of $\simeq$ 60 per \AA\/ are shown by asterisk and open circle
symbols for values away from and in the direction of the PA,
respectively.  Additional measurements for rotation and $\sigma$,
obtained for a lower S/N binning of $\simeq$ 30-35 per \AA, define
greater spatial detail and extend to greater radii; these are plotted
as open star and open square symbols, away from and in the direction of
the PA, respectively. Measurements of rotation and \hth\/ for radii
away from the direction of PA are multiplied by -1 in folding.  For
some galaxies (NGC\,2778, NGC\,4339, NGC\,4464, NGC\,4468, NGC\,4478,
NGC\,5582) the major axis rotation and \hth\/ measurements are plotted
such that the rotation shows mostly positive values.

Where a chromatic focus variation correction for a particular spectrum
was applied, this is indicated and the corresponding effective
smoothing in arcsec is given. Considerable reference is made to
photometry available in the literature, e.g., U, B, and R-band
photometry of Peletier et~al. (1990, hereafter P90), HST WFPC-1 V-band
photometry of van den Bosch et~al. (1994, hereafter vdB94) and Lauer
et~al. (1995, hereafter L95) and HST WFPC-2 V and I-band photometry of
Carollo et~al. (1997, hereafter C97).

Tables for measurements presented in this appendix are available
electronically from the Centre de Donnees astronomiques de Strasbourg
(CDS).

\subsection{NGC\,2778}
Classified as elliptical in RC3, rotation and velocity dispersion were
measured previously for this galaxy by DEFIS83 and F95. Photometry was
obtained by P90 who detected a small amount of diskiness for $r \la
7$\arcsec, and a gradual isophotal twisting between 3\farcs2 and
40\farcs1, from 45\fdg8 to 42\fdg1.

\begin{figure}
\centering
    \epsfig{file=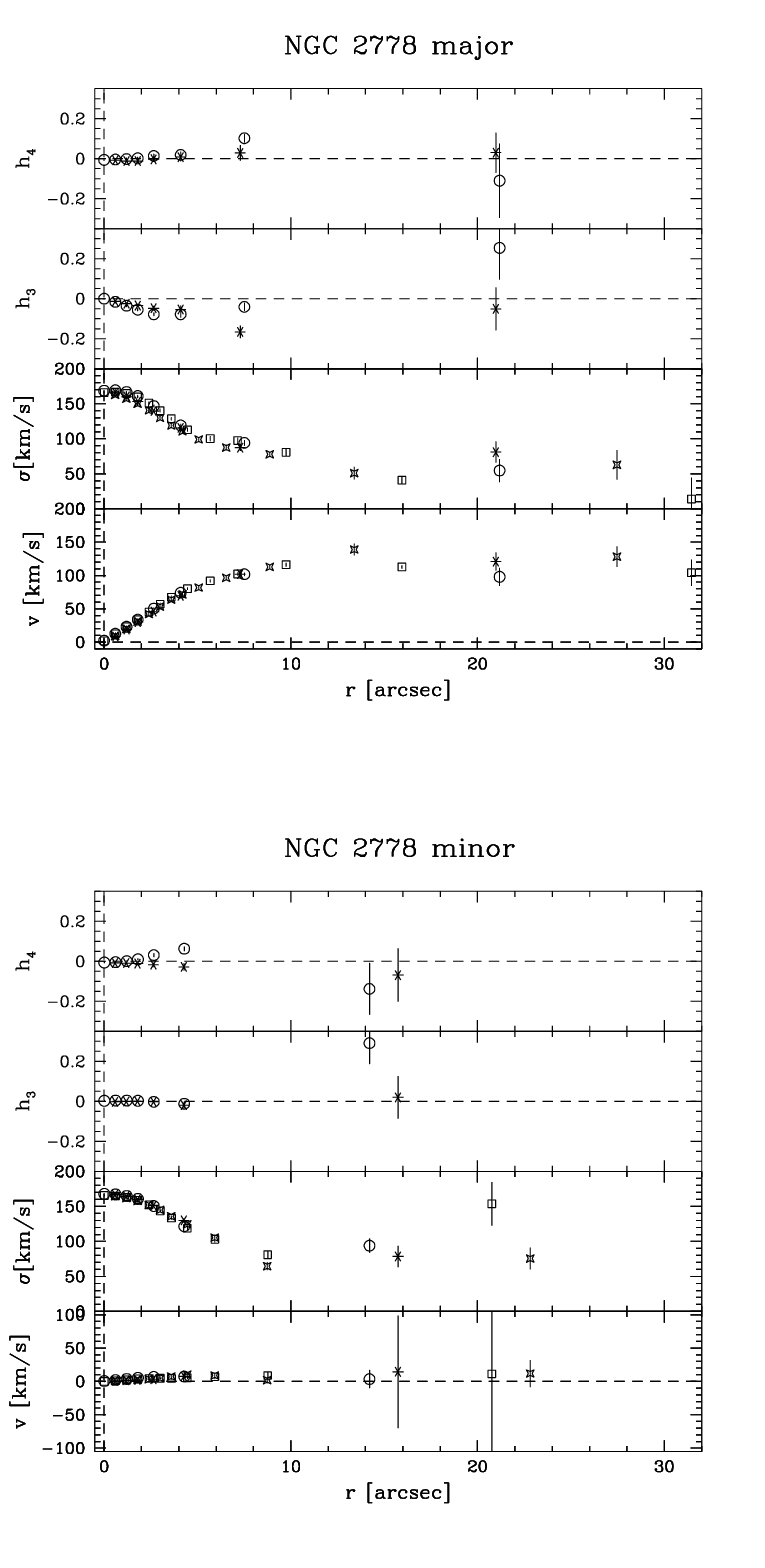,width=8.5cm}
\caption{\small \label{2778mjmn} Major and minor axis observations for
  NGC\,2778 (PA 225\degr\/ and 135\degr, respectively). Focus
  corrections were applied for both spectra corresponding to a maximum
  Gaussian smoothing of 0\farcs81 and 1\farcs08, respectively. The
  seeing for these observations was approximately 1\arcsec. The
  effective seeing, after chromatic focus corrections, is 1\farcs29 and
  1\farcs47, respectively. Data points cannot be assumed to be
  independent at this level. In all figures, measurements obtained for
  a S/N $\simeq$ 60 per \AA\/ are presented by asterisk and open circle
  symbols for values away from the direction of, and in the direction
  of the PA, respectively.  Additional measurements for rotation and
  $\sigma$, obtained for a lower S/N $\simeq$ 30-35 per \AA, are
  plotted as open star and open square symbols, away from and in the
  direction of the PA, respectively.}
\end{figure}

For the major axis, \hth\/ is non-zero for $|r| < 8$\arcsec\/ (see
Figure~\ref{2778mjmn}). \hf\/ is positive for $r \sim 7$\farcs$5$. For
the minor axis, \hth\/ and \hf\/ are mostly consistent with zero; only
at $r \sim 4\arcsec$ is a positive \hf\/ detected.

\subsection{NGC\,3377}
A member of the Leo~I group, NGC\,3377 is widely considered to be a
``disky'' elliptical. It is classified as an intermediate E5 elliptical
galaxy in RC3 and E6 in Sandage and Tammann (1987, hereafter RSA).
Photometric measurements by Carter (1987), P90 and Scorza \& Bender
(1995, hereafter SB95) have shown this galaxy to be ``disky'' for the
range of radii studied here beyond which it becomes ``boxy''.

\begin{figure}
\centering
    \epsfig{file=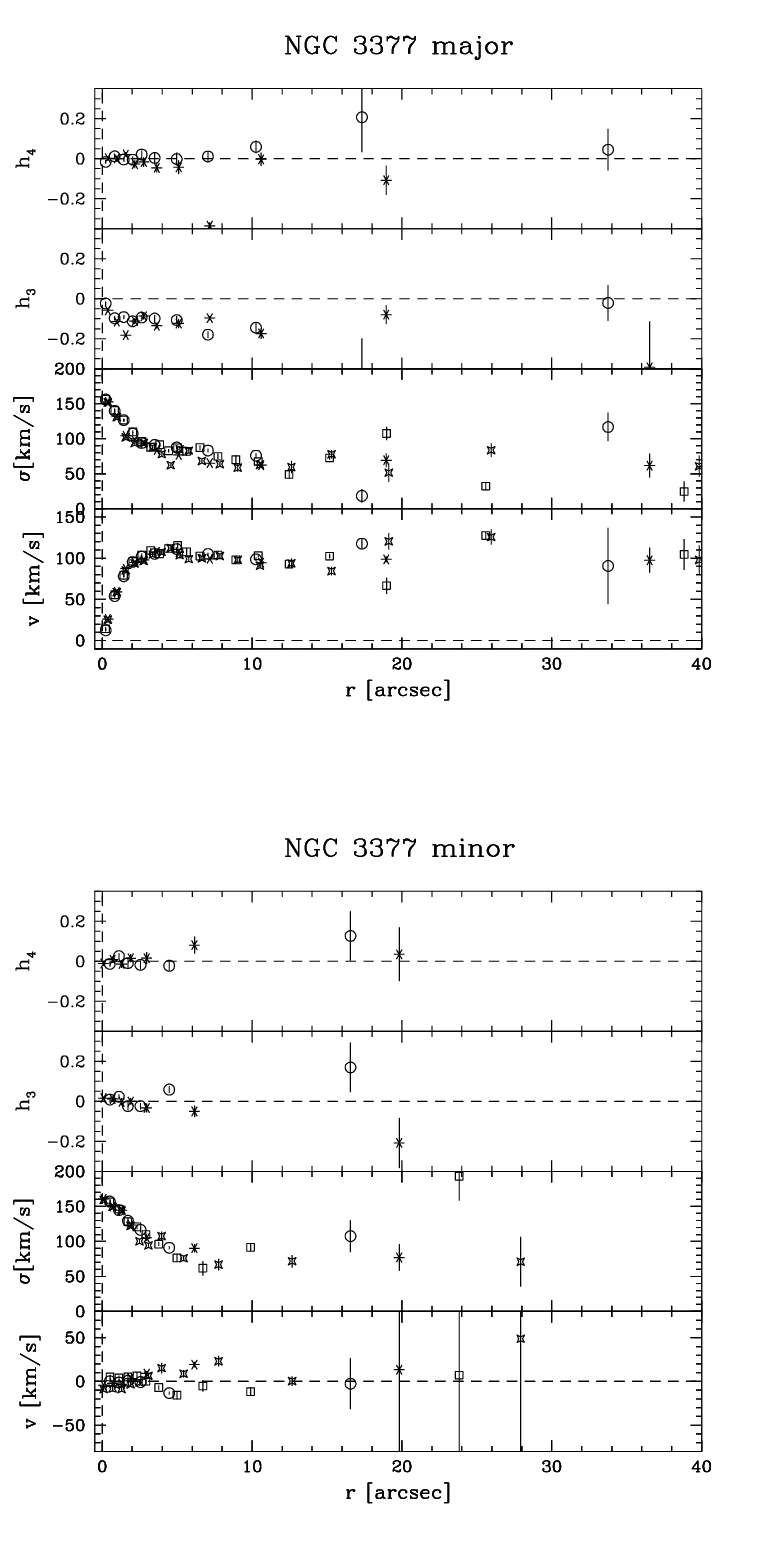,width=8.5cm}
\caption{\small \label{3377mjmn}Major and minor axis observations for
  NGC\,3377 (PA 42\degr\/ and 312\degr, respectively). The seeing for
  both observations was approximately 0\farcs8. }
\end{figure}

For the major axis, \hth\/ is non-zero for almost all radii. \hf\/ is
mostly consistent with zero (see Figure~\ref{3377mjmn}). For the minor
axis, \hth\/ and \hf\/ are mainly consistent with zero. The rotation
curve of the major axis is steeply rising to $\sim$110~\kms\/ within
$r \le 3\arcsec$ and stays roughly constant out to 40\arcsec. The
minor axis shows non-zero velocities for $3\arcsec < |r| < 10
$\arcsec. Our non-zero measurements of \hth\/ along the major axis are
consistent with a central disk at radii $|r| <20\arcsec$ as described
by SB95.

\subsection{NGC\,3379}
NGC\,3379 is a member of the Leo-I group and is a well observed
elliptical, classified as E1 in RC3.

\begin{figure}
\centering
    \epsfig{file=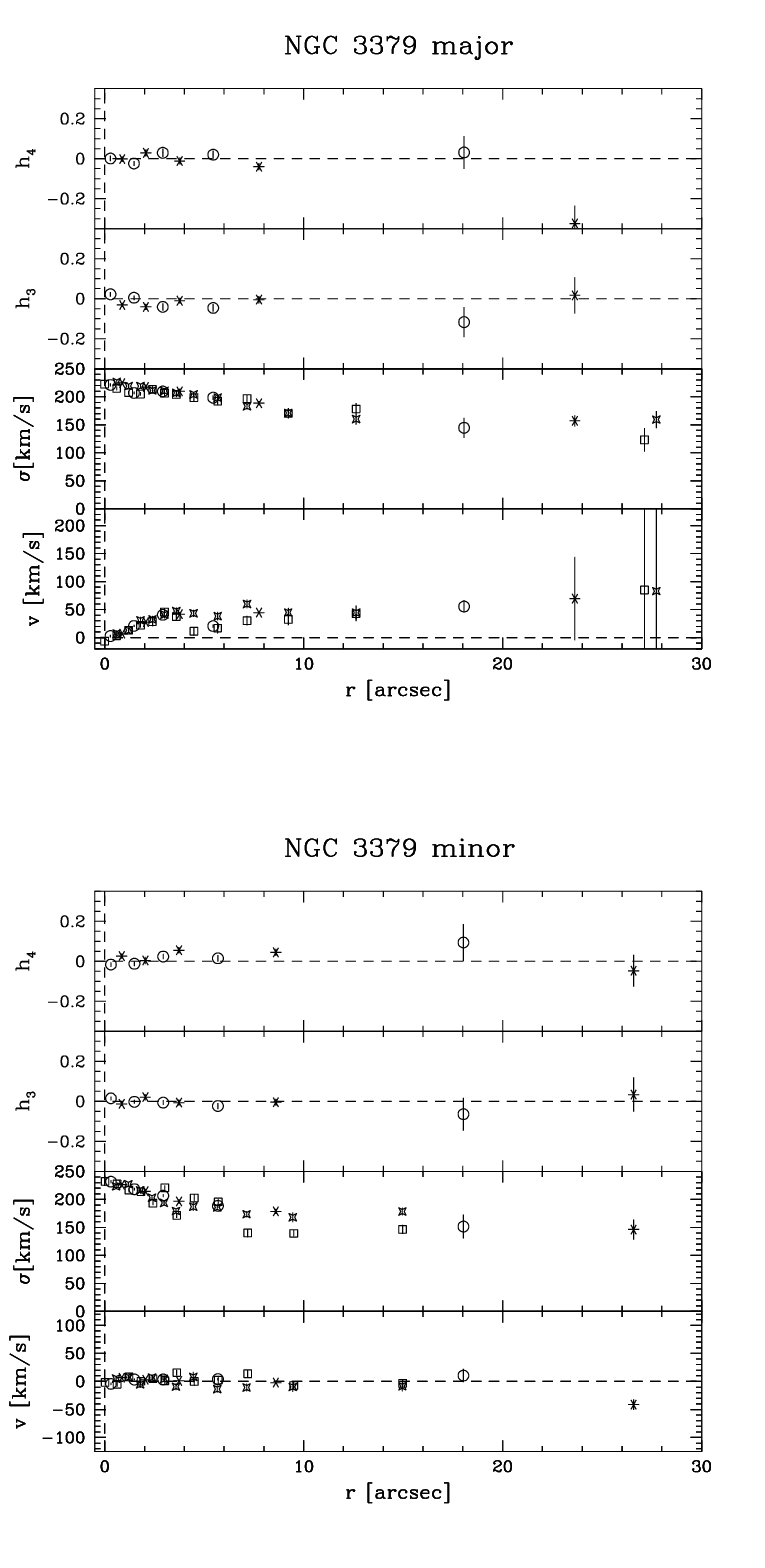,width=8.5cm}
\caption{\small \label{3379mjmn} Major and minor axis observations for
  NGC\,3379 (PA 71\degr\/ and 161\degr, respectively). The seeing for
  both observations was approximately 0\farcs8. }
\end{figure}

For the major axis, rotation decreases at $r \sim 4\arcsec$ and then
increases again at larger radius. Both \hth\/ and \hf\/ show small
non-zero values but there is no clear trend. For the minor axis, both
\hth\/ and \hf\/ are mostly consistent with zero (see Figure~\ref{3379mjmn}).

\subsection{NGC\,3605}
NGC\,3605, a member of the Leo~II group, is classified as an E4
intermediate in RC3, and E5 in RSA. Measurements of rotation and
$\sigma$ for this galaxy were obtained previously by DEFIS83.
Photometry for this galaxy was studied by P90 and L95. For $r \la
4\arcsec$ the galaxy shows ``disky'' isophotes while for $r \ga
4\arcsec$, the isophotes are increasingly ``boxy''.

\begin{figure}
\centering
    \epsfig{file=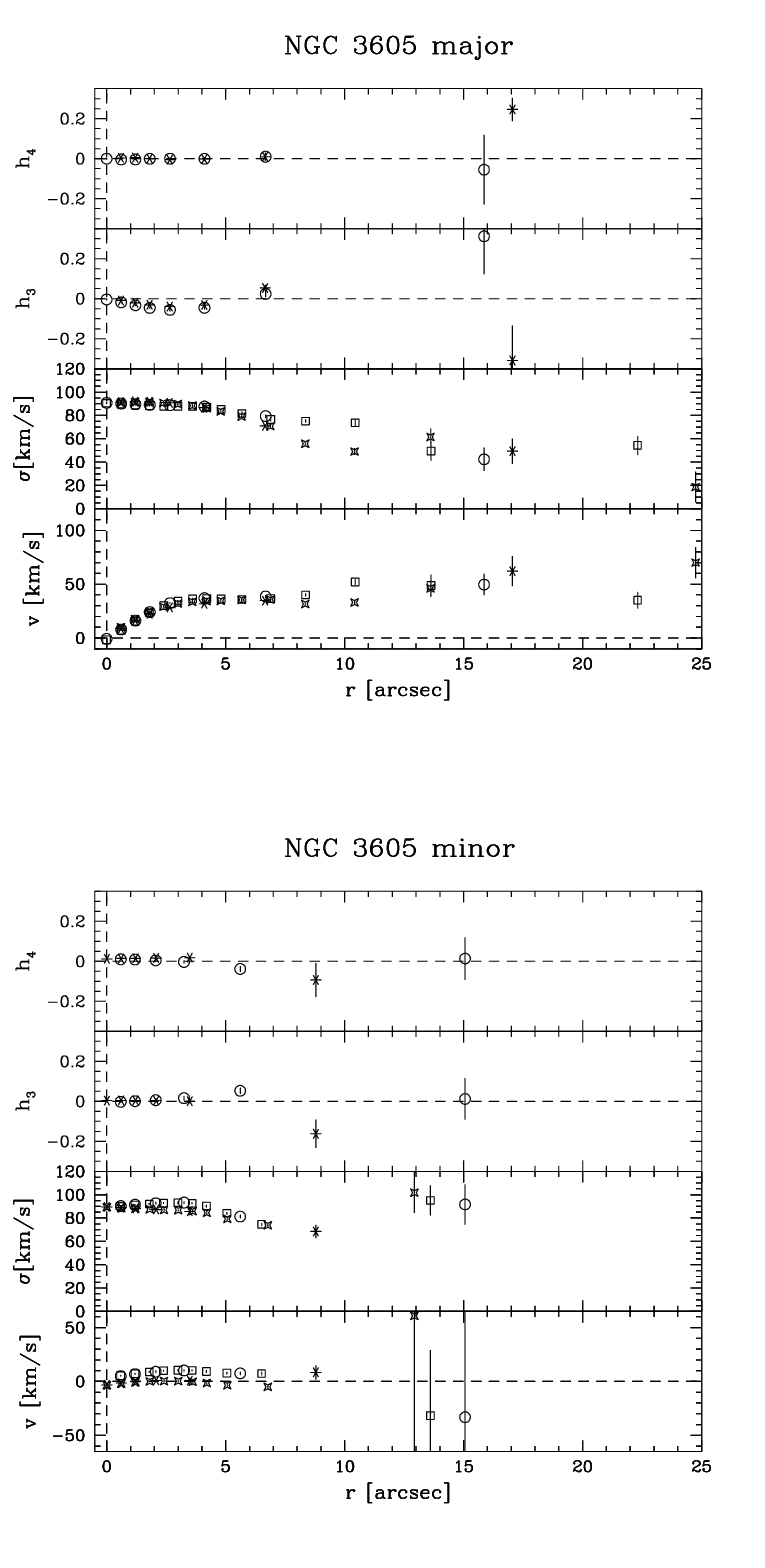,width=8.5cm}
\caption{\small \label{3605mjmn}Major and minor axis observations for
  NGC\,3605 (PA 290\degr\/ and 200\degr, respectively). The minor axis
  observation was subject to significant smoothing during focus
  corrections, i.e. a maximum Gaussian smoothing of 1\farcs44 was
  applied. For the major axis spectrum, a maximum smoothing of
  0\farcs69 was required. The seeing for both observations was
  $\sim$1\arcsec; the total effective smoothing in the spatial
  direction is therefore 1\farcs75 and 1\farcs21 for the minor and
  major axis observations, respectively.}
\end{figure}

For the major axis spectrum, the most notable features are the
non-zero values of \hth\/ for $2\arcsec < |r| < 7\arcsec$ (see
Figure~\ref{3605mjmn}).  \hf\/ is zero for these radii. For the minor
axis, both \hth\/ and \hf\/ are consistent with zero close to the
galaxy centre, becoming non-zero at greater radii. The minor axis also
shows non-zero velocities. It should be noted that the minor axis
spectrum was smoothed considerably in the spatial direction in order
to account for chromatic focus variations. The non-zero measurement of
\hth\/ for the major axis and the detection of diskiness by L95 at
similar radii, support the existence of a central disk component.

\subsection{NGC\,3608}
NGC\,3608, a member of the Leo~II group, is classified as E2 in RC3
and E1 in RSA, this galaxy was one of the first to be recognised to
contain a KDC (JS88). The counterrotation of the galaxy core was
suggested by JS88 to be the result of an interaction with another
galaxy, without full merging, which caused the motion of the outer
parts to be reversed. NGC\,3608 is one of several KDC candidates
imaged by C97. From measurement of the $V-I$ colour gradient, C97
found no significant stellar population differences between the core
and surrounding galaxy of NGC 3608. They found the ellipticity
increased from a value close to zero at $r \simeq 0$\farcs1, to
$\epsilon \simeq 0.2$ at $r \simeq 4 \arcsec$, and NGC\,3608 to be
slightly boxy for radii outside $5\arcsec$.

\begin{figure}
\centering
    \epsfig{file=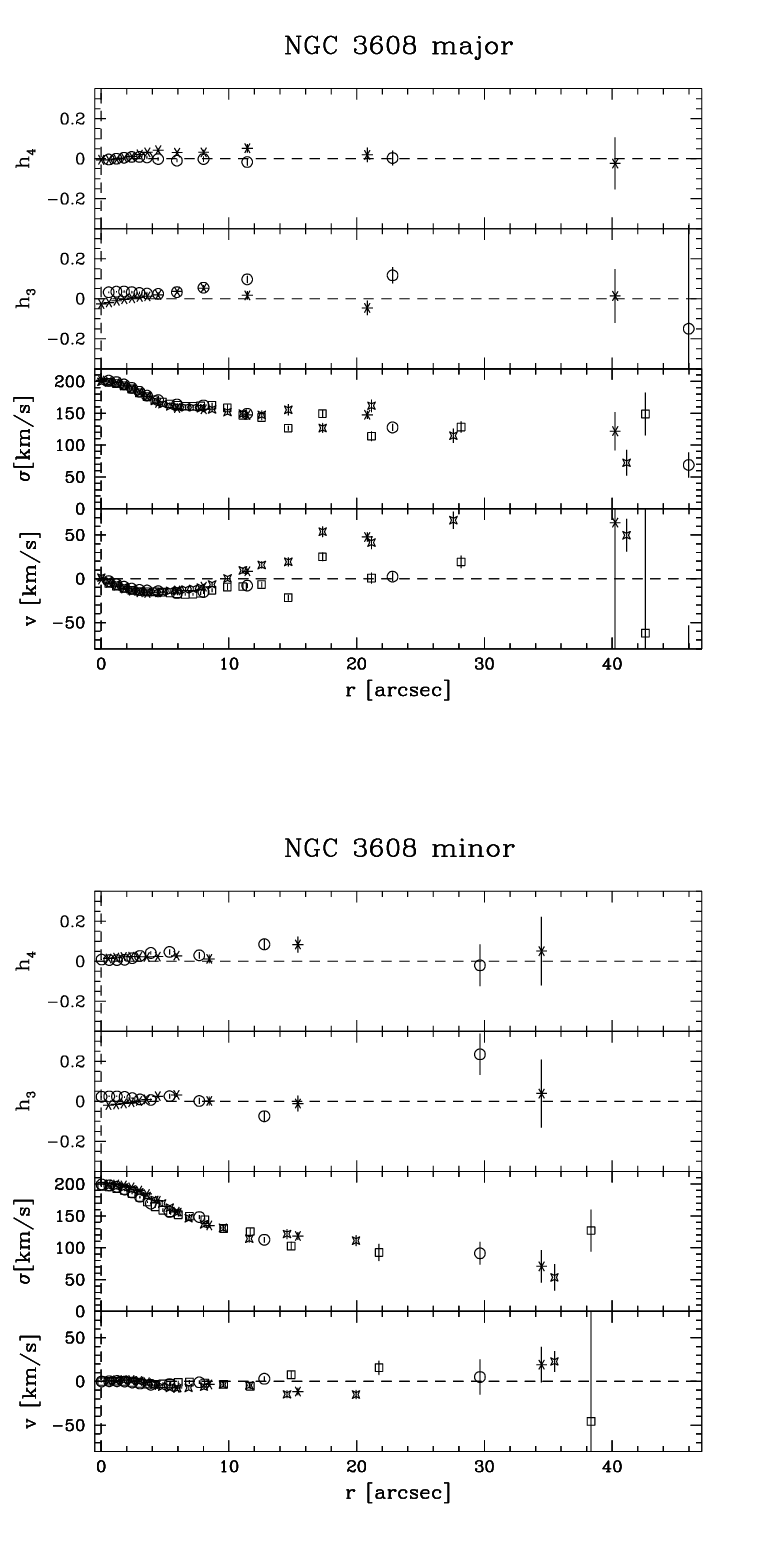,width=8.5cm}
\caption{\small \label{3608mjmn} Measurements for major and minor axes
  of NGC\,3608 (PA 81\degr\/ and 351\degr, respectively). Focus
  corrections were applied for both spectra corresponding to a maximum
  Gaussian smoothing of 0\farcs99 in both cases. The seeing was
  approximately $1\arcsec$ for both spectra. The total smoothing in the
  spatial direction was therefore $\sim$1\farcs5.}
\end{figure}

The profiles of \hth\/ and \hf\/ for both the major and minor axes, are
asymmetric about the galaxy kinematical centre (see
Figure~\ref{3608mjmn}). A small amount of rotation is detected along
the minor axis. The measured velocity along the minor axis for $|r| \ga
10 \arcsec$ is consistent with zero rotation.  Since NGC\,3608 has a
counterrotating KDC, the LOSVD is likely to be more accurately
described by a two-component parameterization.

\subsection{NGC\,4339}
NGC\,4339 is a member of the Virgo cluster and classified as E0 in RC3
and S0${\frac{1}{2}}$ in RSA, this galaxy is intermediate between an
elliptical and S0 galaxy.

\begin{figure}
\centering
    \epsfig{file=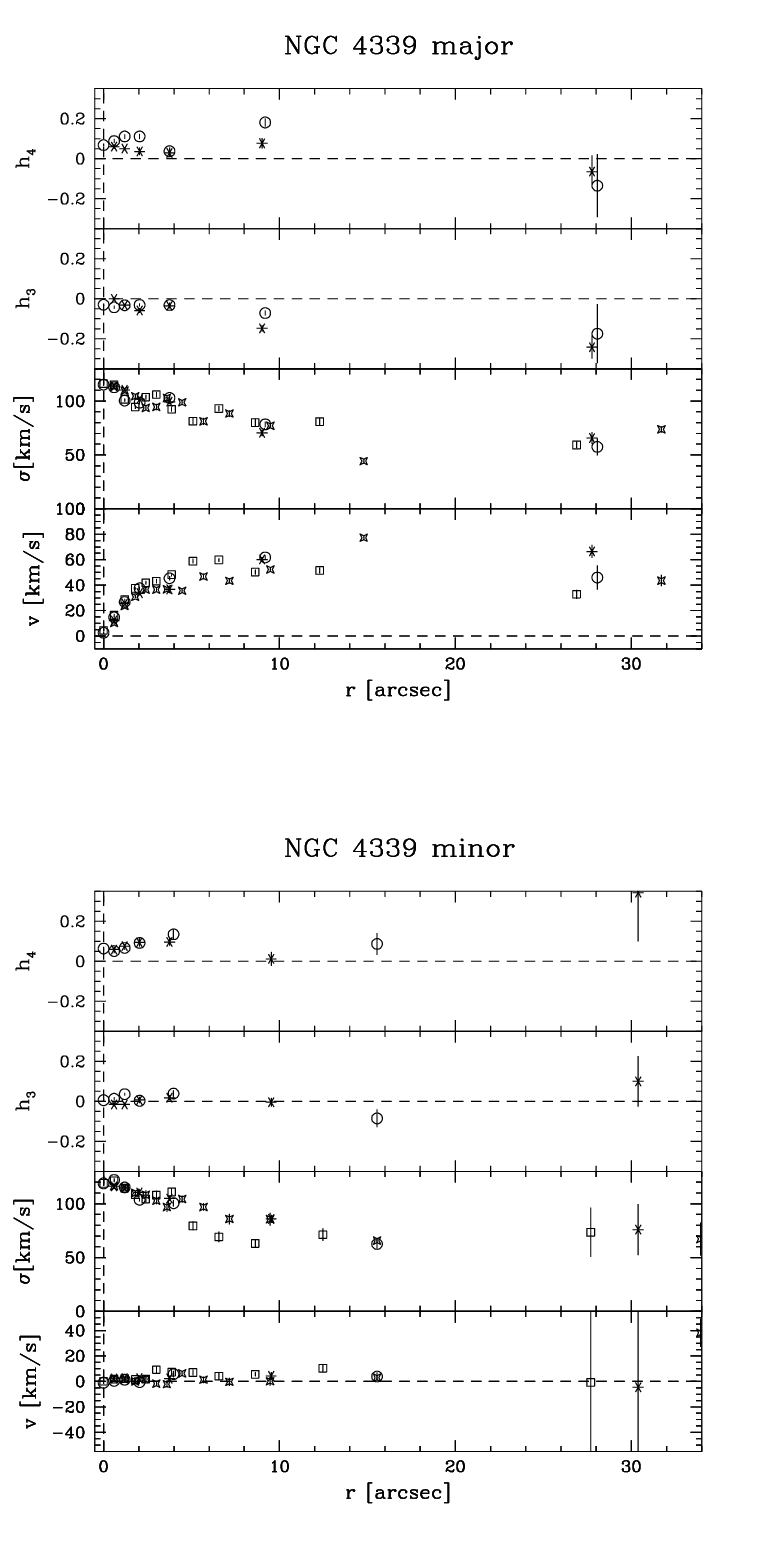,width=8.5cm}
\caption{\small \label{4339mjmn} Major and minor axis observations for
  NGC\,4339 (PA 20\degr\/ and 110\degr, respectively). Focus corrections
  applied to both spectra were small with 0\farcs65 and 0\farcs50,
  respectively. The seeing for both observations was approximately
  0\farcs5. Major axis measurements of rotation have been multiplied by
  -1 for display purposes.}
\end{figure}

For this galaxy, the most striking result is a large positive value of
\hf\/ at almost all radii for both, the major and minor axis. For most
measurements \hth\/ is non-zero along the major axis and consistent
with zero for the minor axis (see Figure~\ref{4339mjmn}).

\subsection{NGC\,4387}
NGC\,4387 is a member of the Virgo cluster and classified as elliptical
in the RC3. This galaxy was one of 8 of the 45 ellipticals imaged by
L95, suspected to contain a nuclear star cluster. Such a component was
proposed to explain an upturn in the measured surface brightness
towards the HST resolution limit.

\begin{figure}
\centering
    \epsfig{file=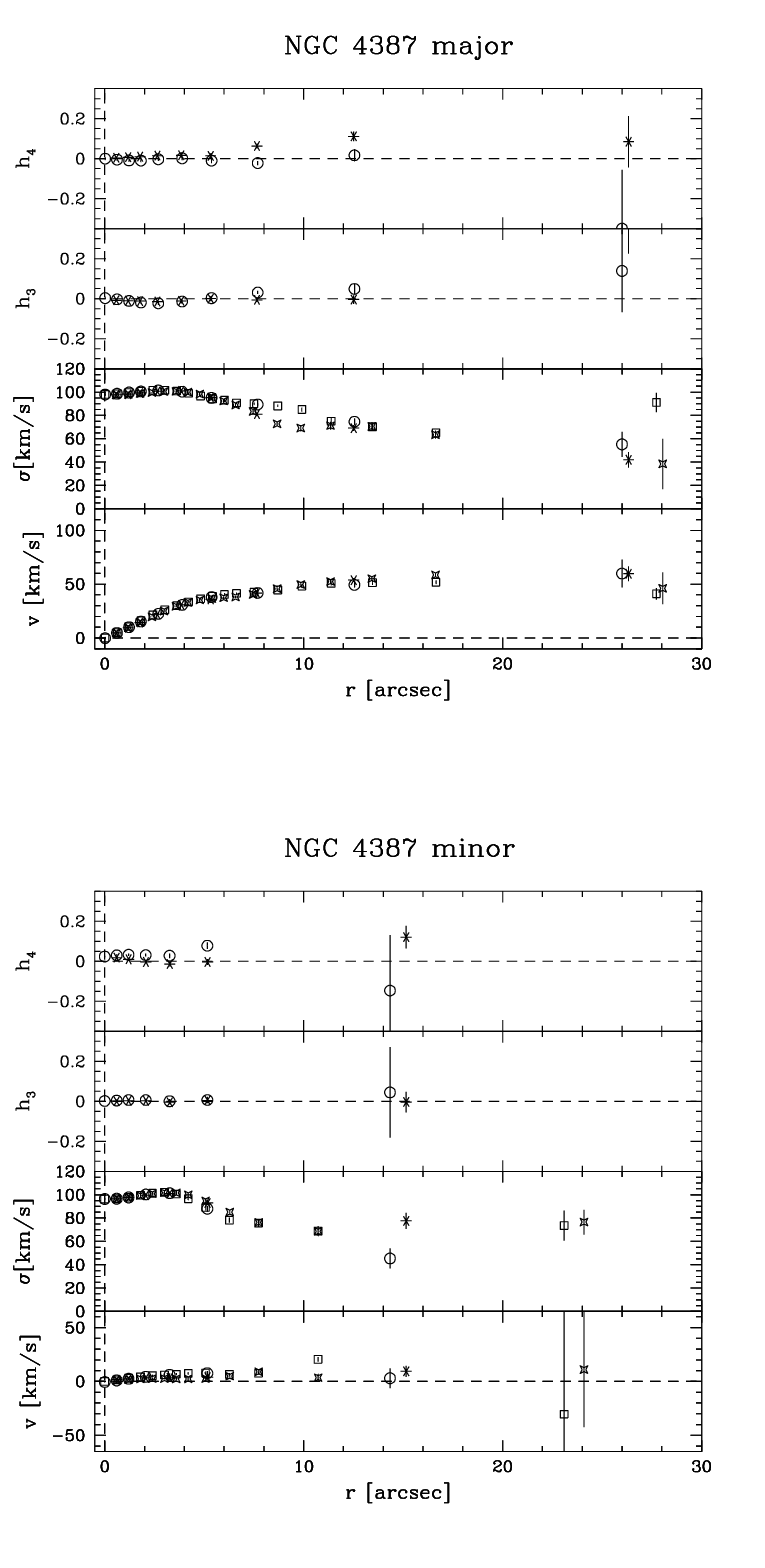,width=8.5cm}
\caption{\small \label{4387mjmn} Major and minor observations for NGC
  4387 (PA 142\degr\/ and 52\degr, respectively). Focus corrections
  corresponding to smoothing of 0\farcs86 and 0\farcs76, respectively,
  were applied. The seeing for both observations was $\sim1\arcsec$;
  the total effective smoothing in the spatial direction is therefore
  1\farcs32 and 1\farcs26. }
\end{figure}

Both major and minor axis observations show a noticeable dip in \csig\/
at the galaxy centre (see Figure~\ref{4387mjmn}). For $0\arcsec \la r
\la 5\arcsec$ along the minor axis, \hf\/ is slightly positive. For the
major axis, \hth\/ is slightly non-zero for $1.5 \arcsec \la |r| \la
5\arcsec$ and positive for $r \ga 7\arcsec$. \hf\/ is positive for $r
\la -5 $\arcsec. Outside $\sim 4 \arcsec$ this galaxy was observed by
P90 to have predominantly boxy isophotes.

The central decrease in $\sigma$ for both axes suggests that a
separate, more rotationally supported component is present at the
galaxy centre.

\subsection{NGC\,4458}
NGC\,4458 is a member of the Virgo cluster and is classified as an E0
intermediate elliptical galaxy in RC3. Studied by L95, its ellipticity
was measured to decrease from a value of $\sim 0.5$ at $r \sim
$0\farcs5 to close to zero at $r \sim $7\farcs5.

\begin{figure}
\centering
    \epsfig{file=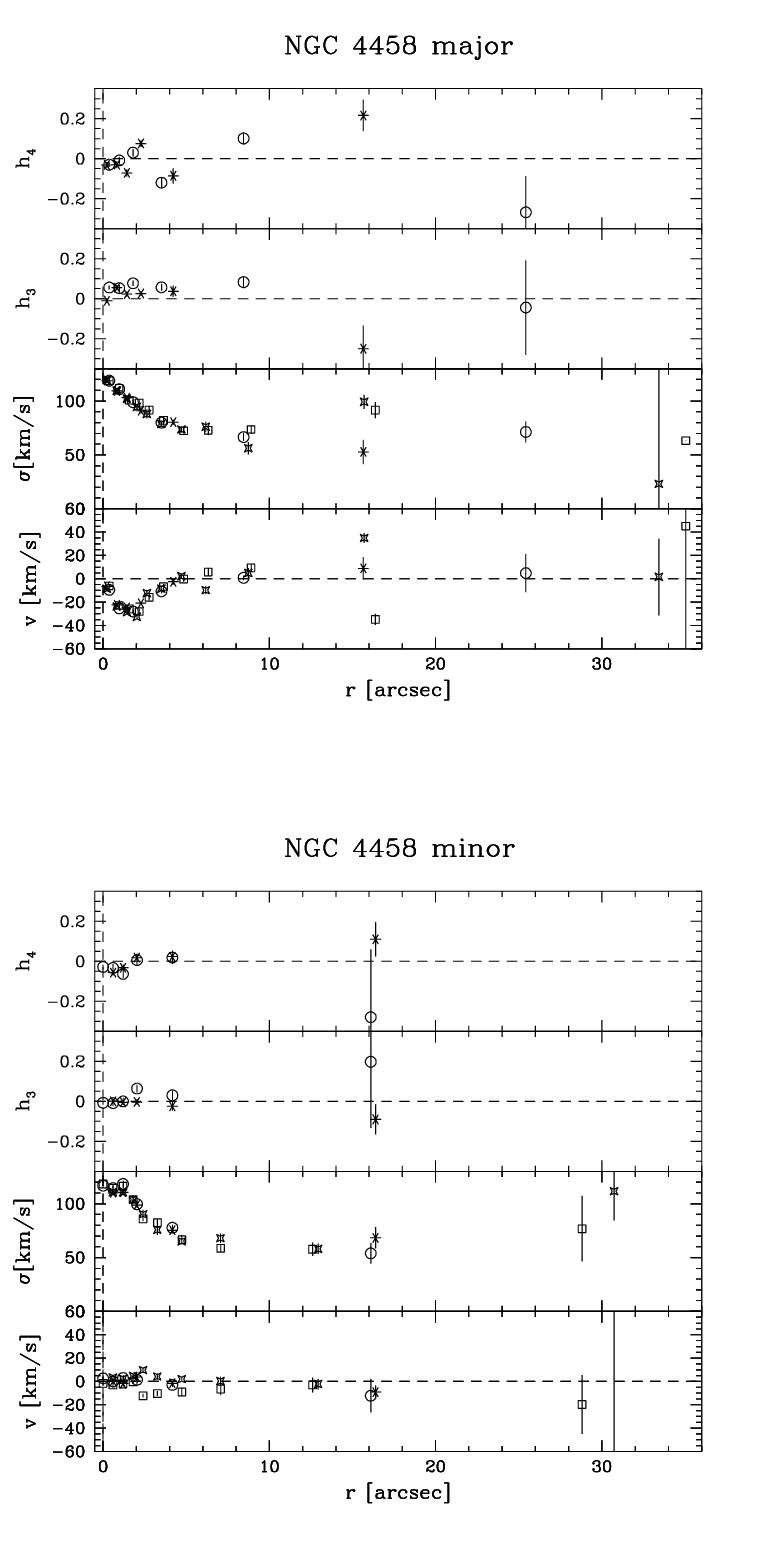,width=8.5cm}
\caption{\small \label{4458mjmn} Major and minor axis observations for
  NGC\,4458 (PA 5\degr\/ and 95\degr, respectively). The seeing for both
  observations was approximately 0\farcs8.}
\end{figure}

Our velocity measurements along the major axis show a clear signature
of a KDC within $|r| < 5\arcsec$ and a maximum rotation speed of
$\sim$30~\kms (see Figure~\ref{4458mjmn}). At radii $|r| >5\arcsec$ we
do not detect any significant rotation. The profiles for both $h_{3}$
and $h_{4}$ are very unusual and change sharply between independent
points. $h_{4}$ is measured to be slightly negative close to the
galaxy centre for both axes. For radii in the direction of the
observation PA, $h_{3}$ is positive for the major axis, and the minor
axis outside the galaxy centre, and is slightly negative for small
negative radii on the major axis.

\subsection{NGC\,4464}
NGC\,4464 is a member of the Virgo cluster and has an uncertain
classification as a spiral galaxy in RC3.  Observed by L95, a$_{4}$ was
measured to be almost zero for the central 4\arcsec, and ellipticity
$\epsilon \sim 0.35$ for $1 \arcsec \la r \la 10 \arcsec$.

\begin{figure}
\centering
    \epsfig{file=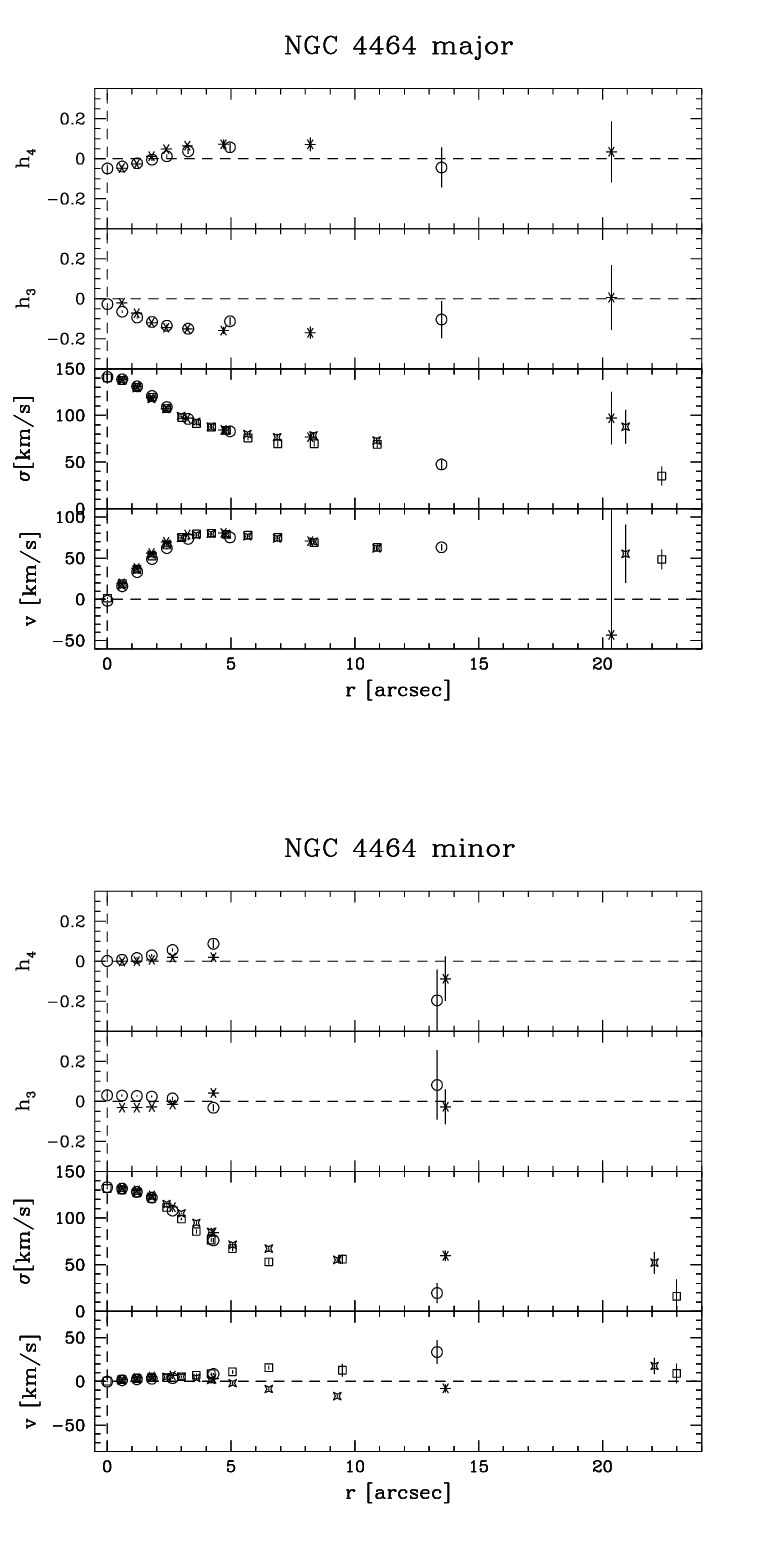,width=8.5cm}
\caption{\small \label{4464mjmn} Major and minor axis observations for
  NGC\,4464 (PA 5\degr\/ and 95\degr, respectively). Focus corrections
  applied to both spectra were 0\farcs61 and 0\farcs75, respectively.
  The seeing for both observations was \csim1\farcs0. Measurements are
  therefore independent only on spatial intervals greater than
  \csim1\farcs25. Major axis measurements of rotation have been
  multiplied by -1 for display purposes.}
\end{figure}

For radii $|r| \la 10 \arcsec$ along the major axis, significant
asymmetrical and symmetrical deviations of the LOSVD from a Gaussian
are indicated by non-zero \hth\/ and \hf\/ (see Figure~\ref{4464mjmn}).
The measurements of \hf\/ are of large amplitude and change sign with
increasing radius. The large amplitude of \hth\/ for the major axis
suggests the presence of multiple components of different kinematics
along the line-of-sight. For $|r| \la 5 $\arcsec, the measurements are
consistent with the superposition of a bulge and an additional, more
rotationally supported, component. At greater radii, the rotation
begins to decrease suggesting that the more rotationally supported
(perhaps disk) component has a scale-length of $\sim 5$\arcsec, and
that at $|r| > 5 $\arcsec\/ a more slowly rotating bulge component
begins to dominate the galaxy light. We note that the kinematic
measurements for this galaxies are similar to those of NGC\,5582.

\subsection{NGC\,4468}
\begin{figure}
\centering
    \epsfig{file=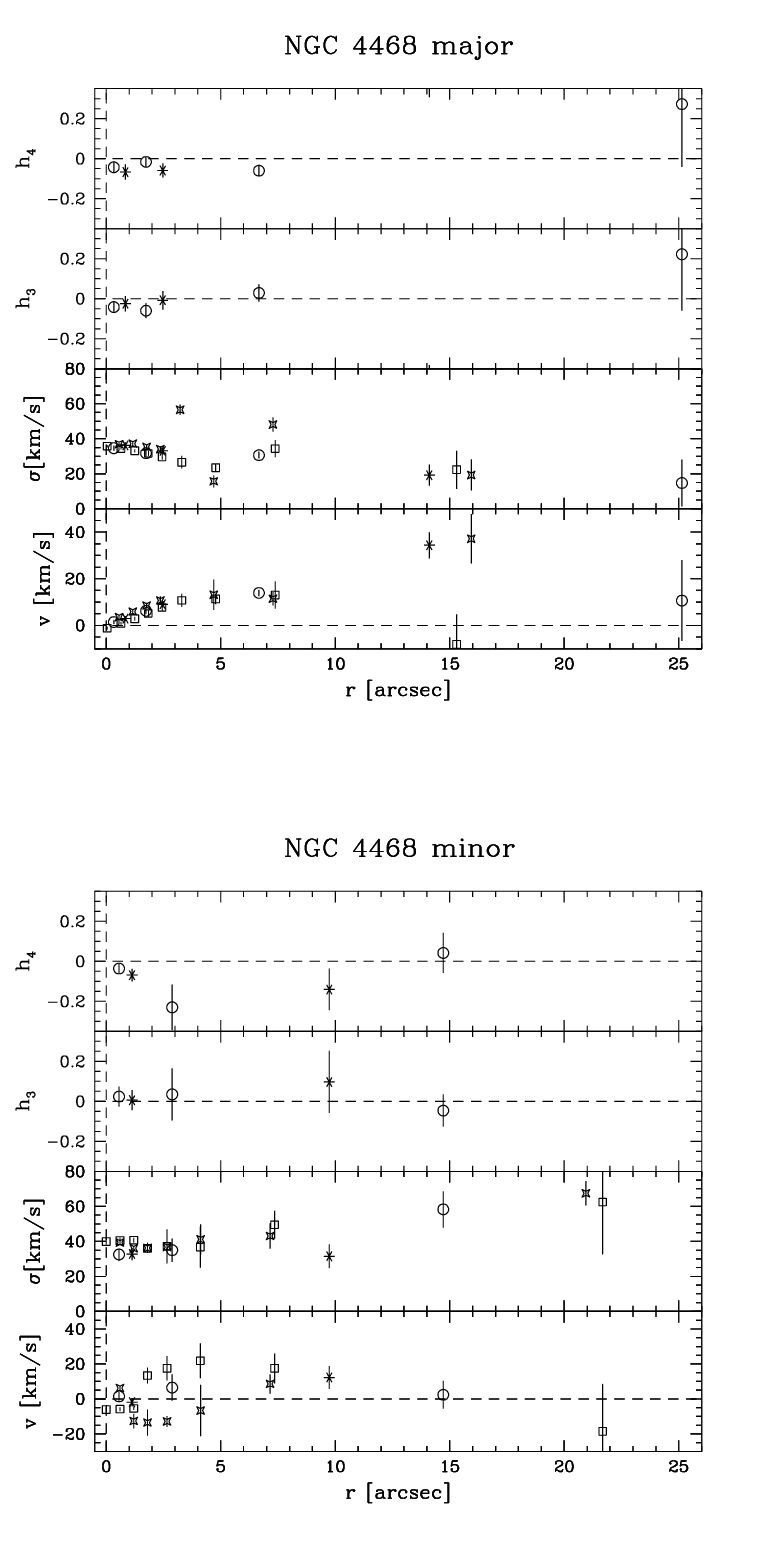,width=8.5cm}
\caption{\small \label{4468mjmn} Major and minor axis observations for
  NGC\,4468 (PA 65\degr\/ and 155\degr, respectively). The seeing for both
  observations was approximately $1\arcsec$. Major axis measurements of
  rotation have been multiplied by -1 for display purposes.}
\end{figure}

NGC\,4468 is a member of the Virgo cluster and has an uncertain
classification as S0 in the RC3. Our exposures for the major and minor
axes of NGC\,4468 were too short to provide good spatial detail for the
LOSVD higher order terms. Taking our errors into account \hth\/ is
almost entirely consistent with zero for both the major and minor axes.
\hf\/ shows negative values for both major and minor axis (see
Figure~\ref{4468mjmn}).

Probably a low surface brightness dwarf elliptical or S0 galaxy, the
measured rotation and velocity dispersion for this galaxy are very low,
with $v$/$\sigma \sim$1.0. These measurements mark the limits
achievable for the spectral resolution of our data.

\subsection{NGC\,4478}
\begin{figure}
\centering
\epsfig{file=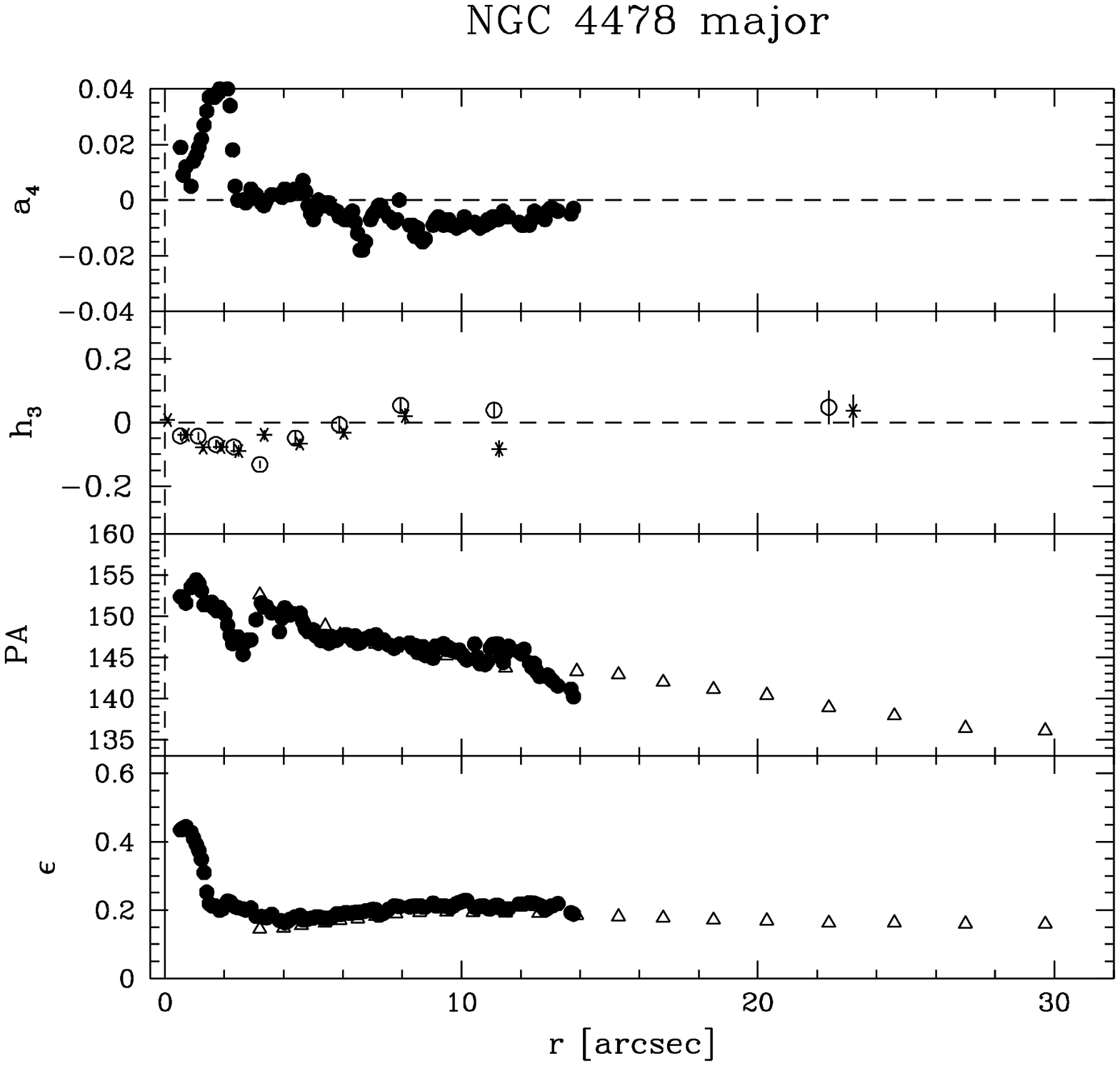,width=8.5cm}
\caption{\small \label{cfkinph4478} Our measurements of \hth\/ 
  for the major axis of NGC\,4478, are shown along with photometric
  measurements of $\epsilon$, PA and a$_4$ from the literature; vdB94
  filled circles, P90 open triangles. Measurements of \hth\/ have been
  folded by multiplying measurements for negative values of radius by
  -1.}
\end{figure}

NGC\,4478 is a member of the Virgo cluster. While classified as E2 in
both RC3 and RSA, this elliptical is classified as a compact
elliptical by Prugniel et~al. (1987). They argue that NGC\,4478 had
been tidally-truncated by its nearby neighbour M87. NGC\,4478 has been
studied with HST by vdB94. In Figure \ref{cfkinph4478} we show their
measurements along the major axis of $\epsilon$, PA and a$_4$ (kindly
supplied by F.C. van den Bosch). Also shown are the measurements of
$\epsilon$ and PA taken from P90.

\begin{figure}
\centering
    \epsfig{file=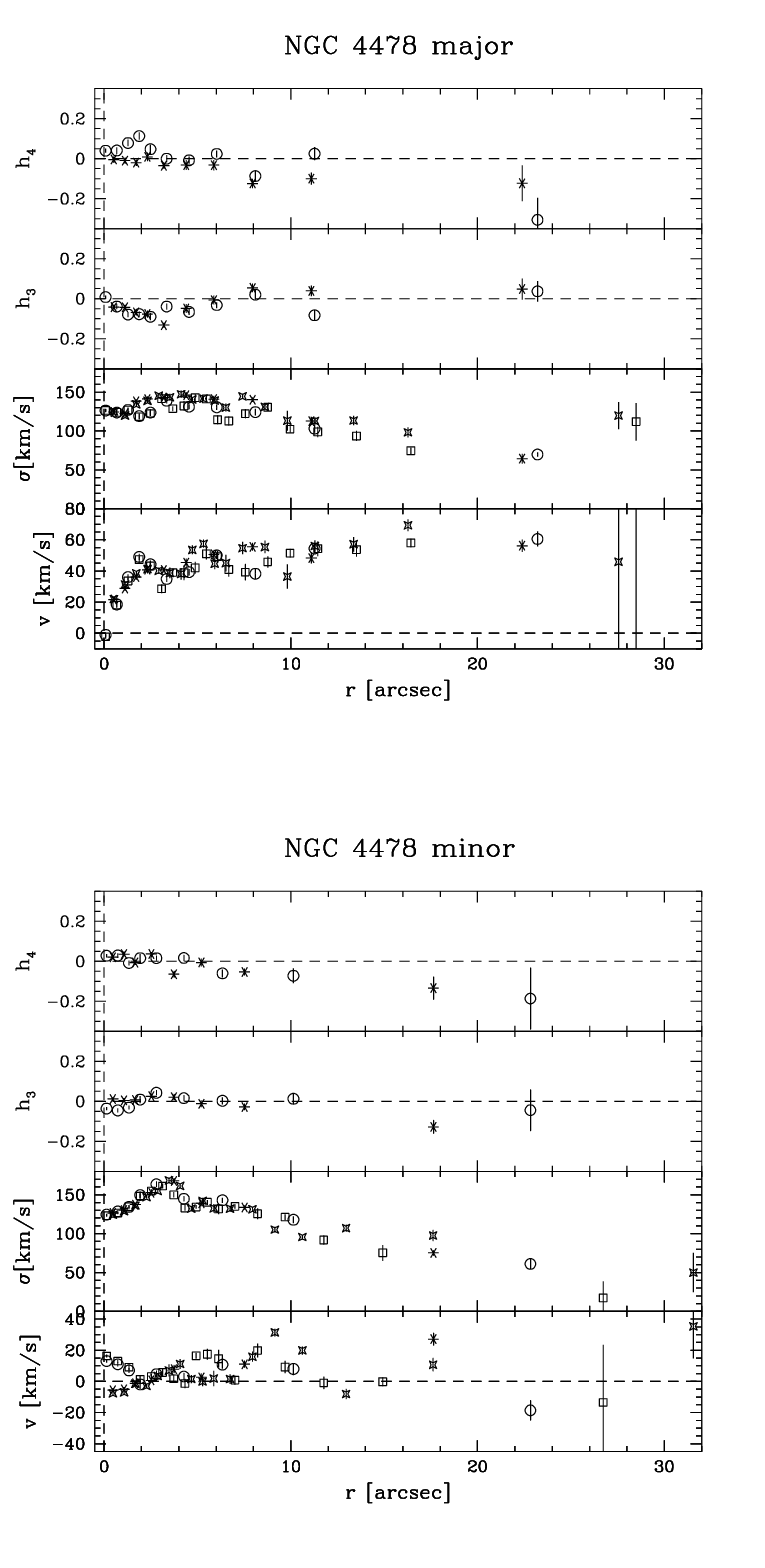,width=8.5cm}
\caption{\small  \label{4478mjmn} Major and minor axis observations of
  NGC\,4478 (PA 145\degr\/ and 55\degr, respectively). The seeing for
  observations was approximately 0\farcs8 and 0\farcs5,
  respectively. Major axis measurements of rotation have been
  multiplied by -1 for display purposes.}
\end{figure}

In Figure \ref{4478mjmn}, measurements of rotation, $\sigma$, \hth\/
and \hf\/ for $|r| < 10 \arcsec$, provide strong indications of a
central component, decoupled from the surrounding galaxy. Measurements
of $\sigma$ and \hf\/ for this range of radii, are asymmetric, and
rotation and \hth\/ are not point-symmetric about the galaxy centre,
most notably along the major axis. This could be the result of
considerable isophote twisting measured by vdB94 for $|r| < 5 \arcsec$,
such that the measured PA deviated by up to 10\degr\/ from that of our
major axis observation (i.e., PA 145\degr) (see Figure
\ref{cfkinph4478}). Observations obtained for the photometric axes of
the main galaxy may not therefore coincide with the similar axes of a
distinct component. A central decrease in $\sigma$ is measured for
both, minor and major axes. For the major axis, a significantly
non-zero value of \hth\/ is measured for $|r| < 5 $\arcsec, and \hf\/
is positive for $0 \la r \la 3 $\arcsec. For the minor axis, \hth\/ is
slightly non-zero for $|r| < 2\arcsec$, and \hf\/ becomes slightly
negative outside $r = 3\arcsec$.

Two additional interesting features of the photometry of vdB94
(Figure~\ref{cfkinph4478}) are the significant measurement of diskiness
for the inner $2\arcsec$\/ and the sharp increase in ellipticity for
$|r|\la1.5\arcsec$. These two results offer strong support to the idea
that a disk component resides at the galaxy centre.  From the study of
residual maps of their photometry, vdB94 were unable, however, to state
conclusively that such a component was observed.

The asymmetric LOSVDs along the major axis (i.e., non-zero measurement
of $h_{3}$) together with the positive measurement of \hf\/ for both
axes close to the galaxy centre, are consistent with the detection of a
separate kinematical component with orbital motion across the
line-of-sight. The central dip in $\sigma$, for similar radii, suggests
that this component is more rotationally-supported than the surrounding
bulge component. Taken together these results may be consistent with
the detection of a central disk population.

The kinematics of this galaxy are clearly complex. Two-dimensional
maps of the velocity field, available using integral field unit
spectroscopy, would greatly enhance the study of its dynamics.

\subsection{NGC\,4551}
\begin{figure}
\centering
    \epsfig{file=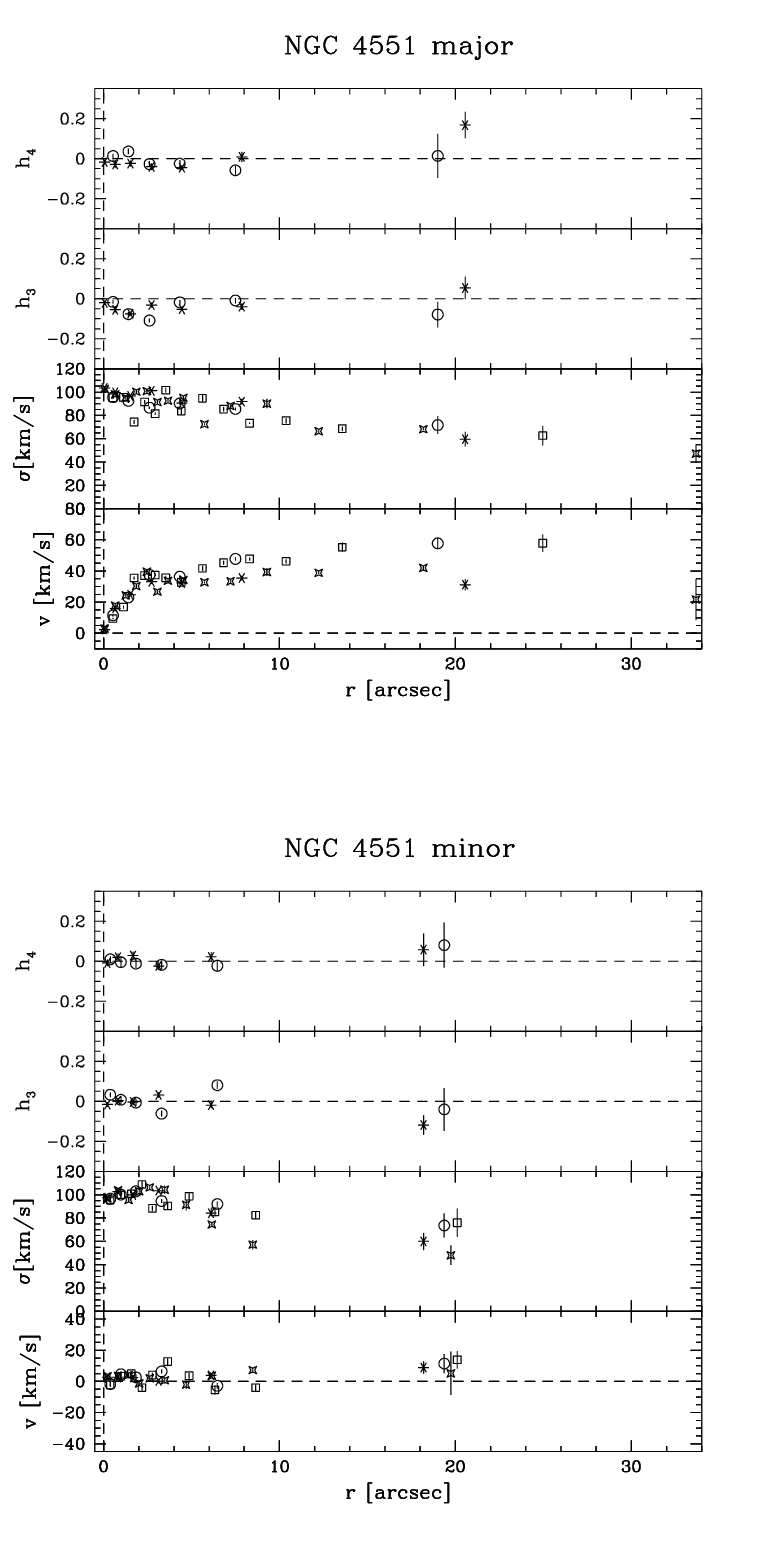,width=8.5cm}
\caption{\small \label{4551mjmn} Major and minor axis observations for
  NGC\,4551 (PA 70\degr\/ and 160\degr, respectively). The seeing for
  both observations was approximately 0\farcs8.}
\end{figure}

NGC\,4551 is a member of the Virgo cluster and has an uncertain
classification as elliptical in RC3. Our kinematic measurements are
presented in Figure \ref{4551mjmn}. This galaxy was studied by both,
P90 and L95. The photometric measurements of L95 found NGC\,4551 to
have ``disky'' isophotes for $1 \arcsec \la |r| \la 4 \arcsec$. As for
NGC\,4387, L95 have postulated that NGC\,4551 may contain a central
stellar nucleus.  P90 measure NGC\,4551 to become increasingly ``boxy''
from $r \sim 4 \arcsec$ to $r \sim 10 \arcsec$, and at greater radii
a$_{4}$ tends to zero.

For the major axis \hth\/ is non-zero for $|r| \la 5 \arcsec$. \hf\/
shows non-zero values close to the galaxy centre.  For the minor axis
\hf\/ changes only slightly with radius and is almost consistent with a
value of zero.

Rotation is measured for the major axis and no rotation is detected for
the minor axis. A small central decrease in $\sigma$ is clearly
observed for the minor axis. For the major axis this effect is less
pronounced. The dip in central velocity dispersion and the measured
LOSVD asymmetry for $|r| \la 5\arcsec$ along the major axis may be
consistent with the detection of a distinct central component of
orbital motion across the line-of-sight.

\subsection{NGC\,4564}
\begin{figure}
\centering
    \epsfig{file=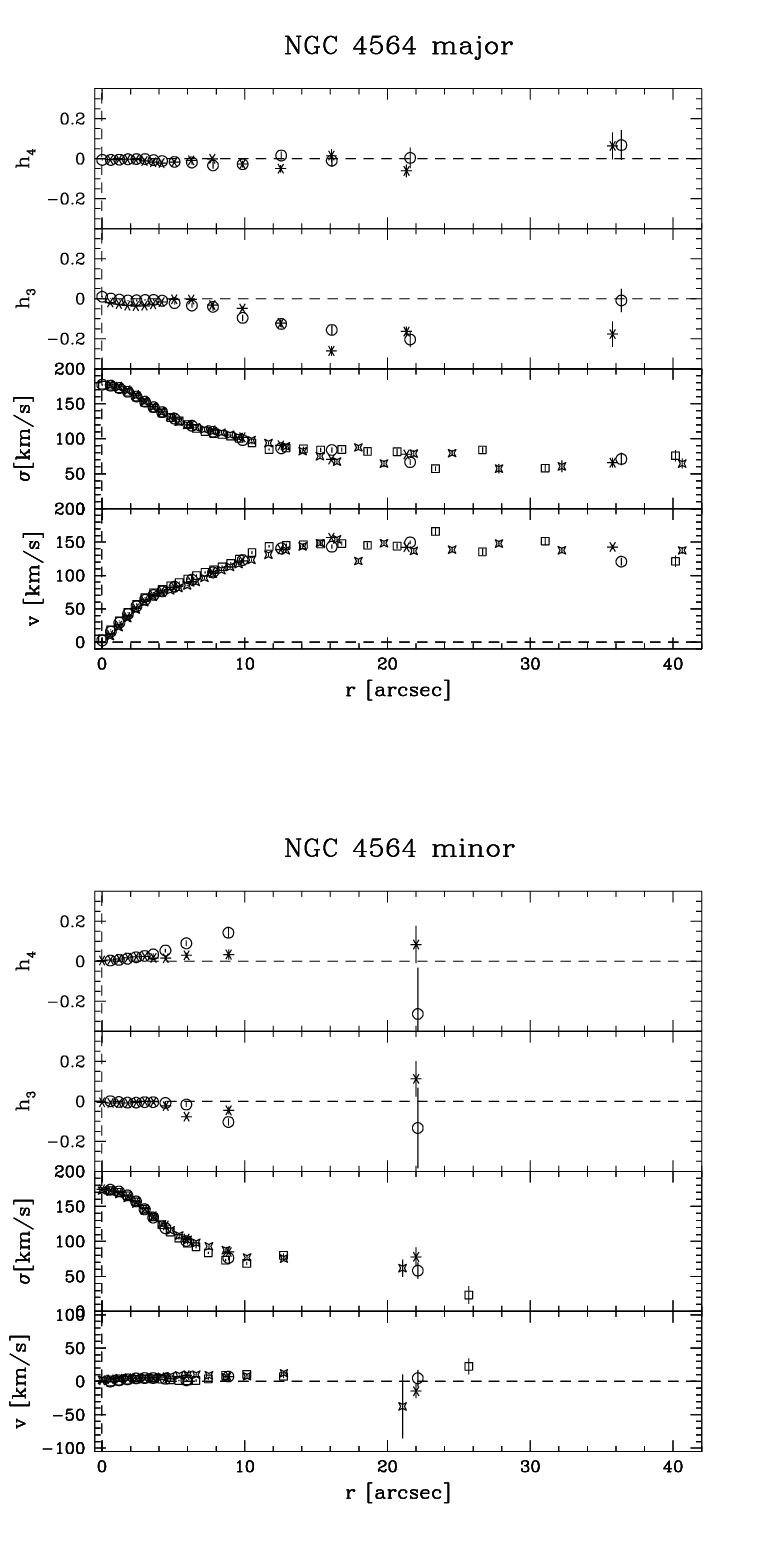,width=8.5cm}
\caption{\small \label{4564mjmn}Measurements for the major and minor
  axes of NGC\,4564 (PA 45\degr\/ and 135\degr, respectively). Both
  spectra were corrected for the effects of focus variations; the
  maximum smoothing applied, 0\farcs82 for the major axis spectrum, and
  0\farcs74 for the minor axis spectrum, was in both cases smaller then
  the seeing of $\sim 1$\arcsec. The total effective smoothing is
  1\farcs29 and 1\farcs24, respectively.}
\end{figure}

\begin{figure}
\centering
\epsfig{file=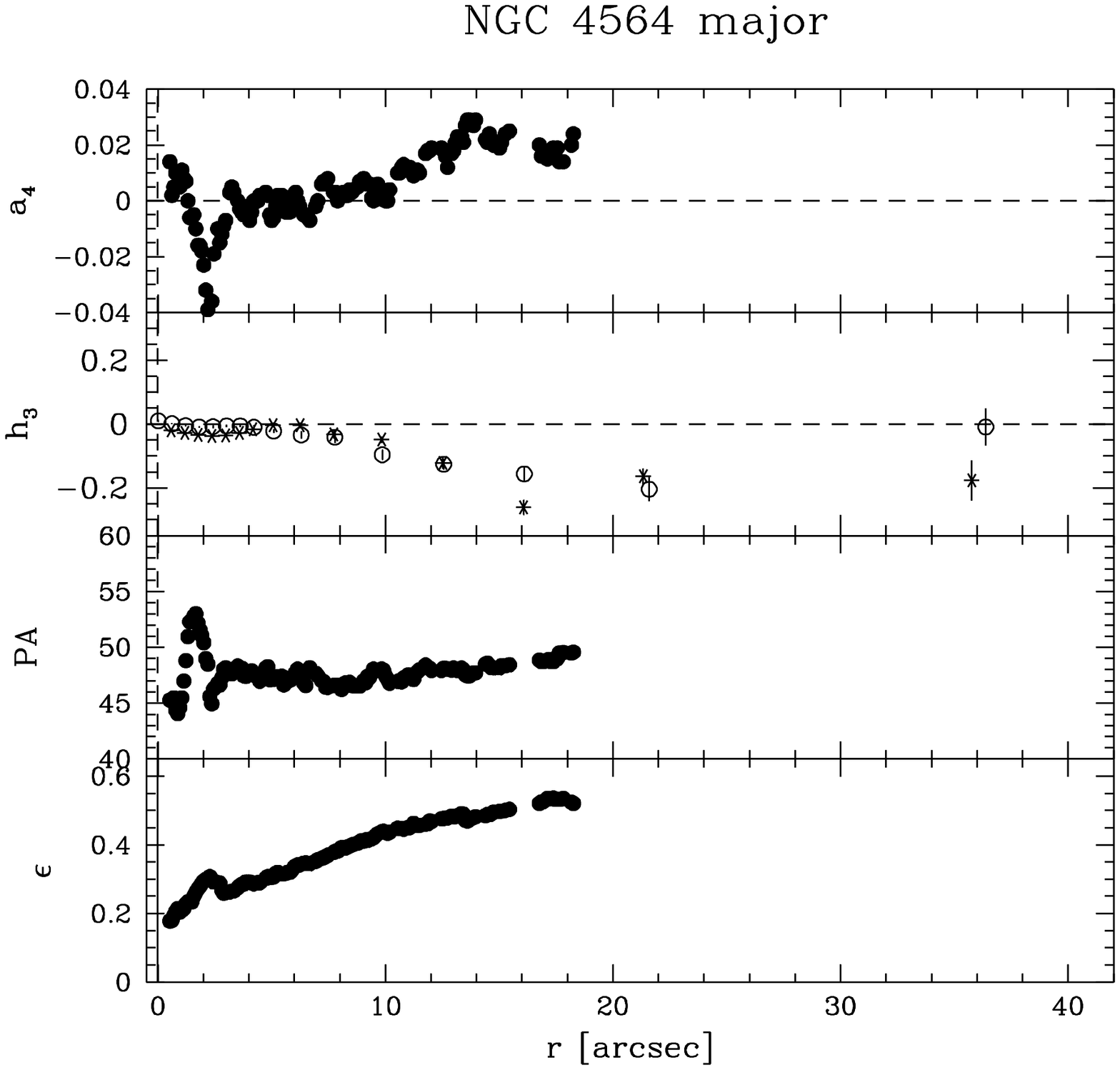,width=8.5cm}
\caption{\small \label{cfkinph4564} h$_3$ measurements for the major 
  axis of NGC\,4564 obtained here are plotted together with the
  photometric measurements of van den Bosch et~al. (1994) (i.e.
  measurements of a$_4$, ellipticity ($\epsilon$) and PA as functions
  of radius). Measurements of \hth\/ have been folded by multiplying
  measurements for negative values of radius by -1. }
\end{figure}

NGC\,4564 is a member of the Virgo cluster and classified as E6 in both
RC3 and RSA. Measurements of v, $\sigma$, \hth\/ and \hf\/ are given in
Figure \ref{4564mjmn}. This galaxy has been studied by vdB94 using HST
photometry and in Figure \ref{cfkinph4564} major axis measurements of
\hth\/ obtained here are compared with the photometric measurements of
$\epsilon$, PA and a$_4$ from vdB94.

Non-symmetric LOSVDs are detected along the major axis by non-zero
\hth\/ for $|r| \ga 5$\arcsec. \hf\/ is consistent with zero. For the
minor axis \hth\/ is non-zero for $-9 $\arcsec$ \la r \la -5\arcsec$
and $r \sim 9$\arcsec. \hf\/ is positive for $|r| \ge 5\arcsec$.

In Figure \ref{cfkinph4564}, the most interesting results from the
photometry are the measurement of negative a$_4$ and hence boxiness for
$r \sim 2\arcsec$, and measurement of diskiness for $|r| \ga
10$\arcsec, isophote twisting for $r < 2$\farcs5, and gradually
increasing measurement of \ell\/ with radius. The measurements of a$_4$
for different radii suggest that two separate components are being
detected: within the inner 6\arcsec\/ the measurements of boxy
isophotes are consistent with existence of a central bulge; diskiness
for $r \ga 10 \arcsec$, where rotation is still measured to increase
with radius, and the gradual increase in \ell\/ with radius, are
consistent with the detection of a disk component. Measurements are
consistent with detection of a disk component embedded within a more
slowly rotating bulge.

\subsection{NGC\,5582}
\begin{figure}
  \centering
    \epsfig{file=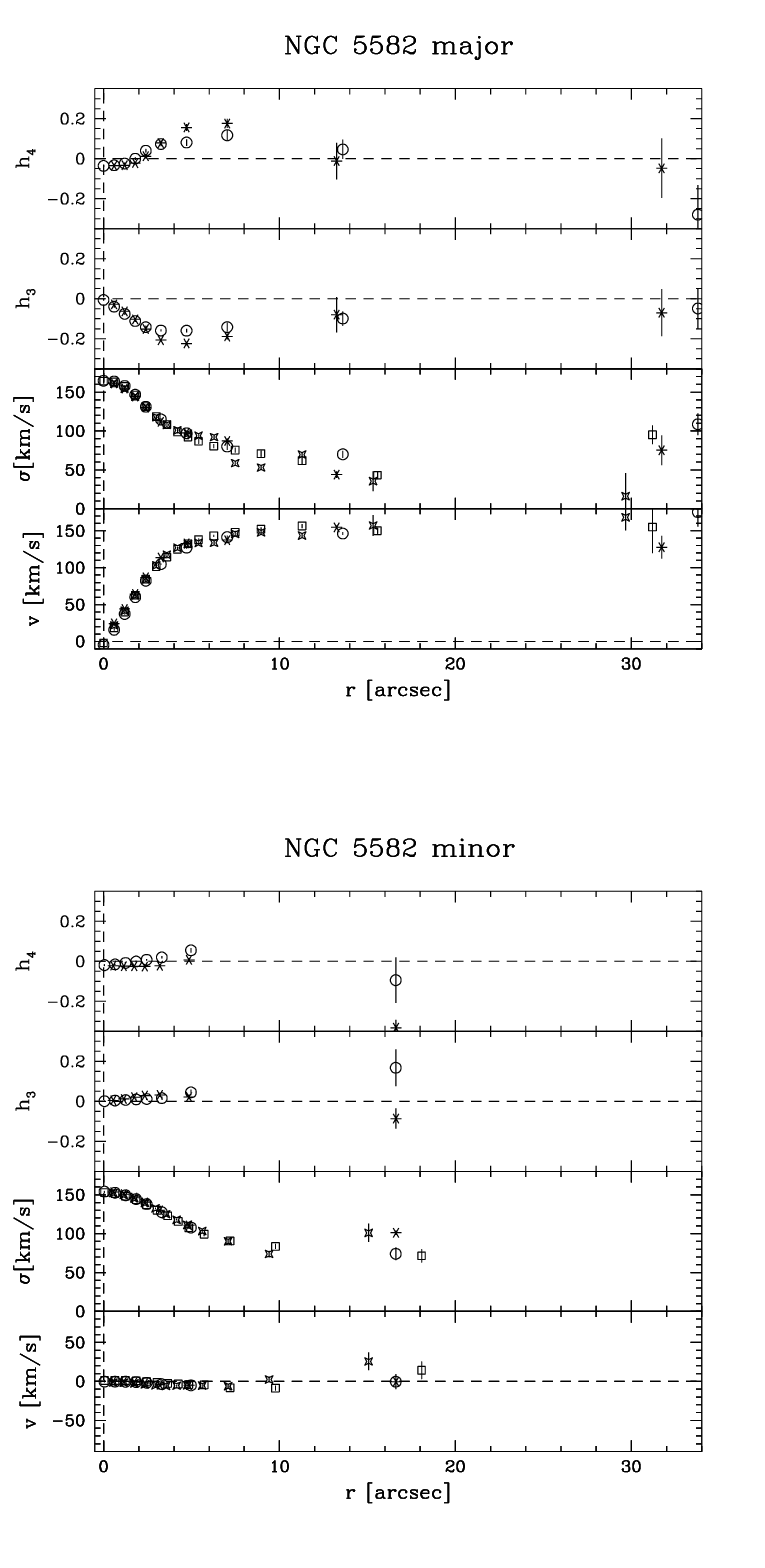,width=8.5cm}
\caption{\small \label{5582mjmn} Major and minor axis observations for
  NGC\,5582 (PA 25\degr\/ and 115\degr, respectively). Focus
  corrections applied for the major and minor axis spectra corresponded
  to a {\it maximum} Gaussian smoothing of 0\farcs74 and 1\farcs30,
  respectively. The seeing for the minor axis observation was 1\arcsec;
  measurements for this spectrum can not be assumed to be independent
  on spatial intervals \protect\linebreak \protect$\simeq$ 1\farcs 6.}
\end{figure}

NGC\,5582 is classified as elliptical in RC3. Measurements are
presented in Figure \ref{5582mjmn}. For the major axis, \hth\/ is
significantly non-zero for $|r| < 14 \arcsec$ and reaches a maximum
amplitude for radii at which there is a knee in the rotation curve.
\hf\/ changes significantly in value with radius. For the inner
10\arcsec, the measurements closely resemble those for the major axis
of NGC\,4464.

Examples of actual fits to the recovered galaxy broadening function,
i.e. the recovered LOSVD shape, are given in Figure~\ref{5582mj_BF} in
the main part of this paper. For $r < -1$\arcsec, the LOSVD is clearly
asymmetric and for $r < -2 \arcsec$ there is clear evidence for the
existence of two separate kinematical components, one of which has
greater mean velocity along the line-of-sight and appears to dominate
the measured kinematics. This is consistent with detection of a
separate bright disk component, or a bulge component of high rotational
support.

\section{Comparisons with other authors}\label{cfother}
\begin{figure*}
\begin{center}
\begin{minipage}{6in}
\epsfig{file=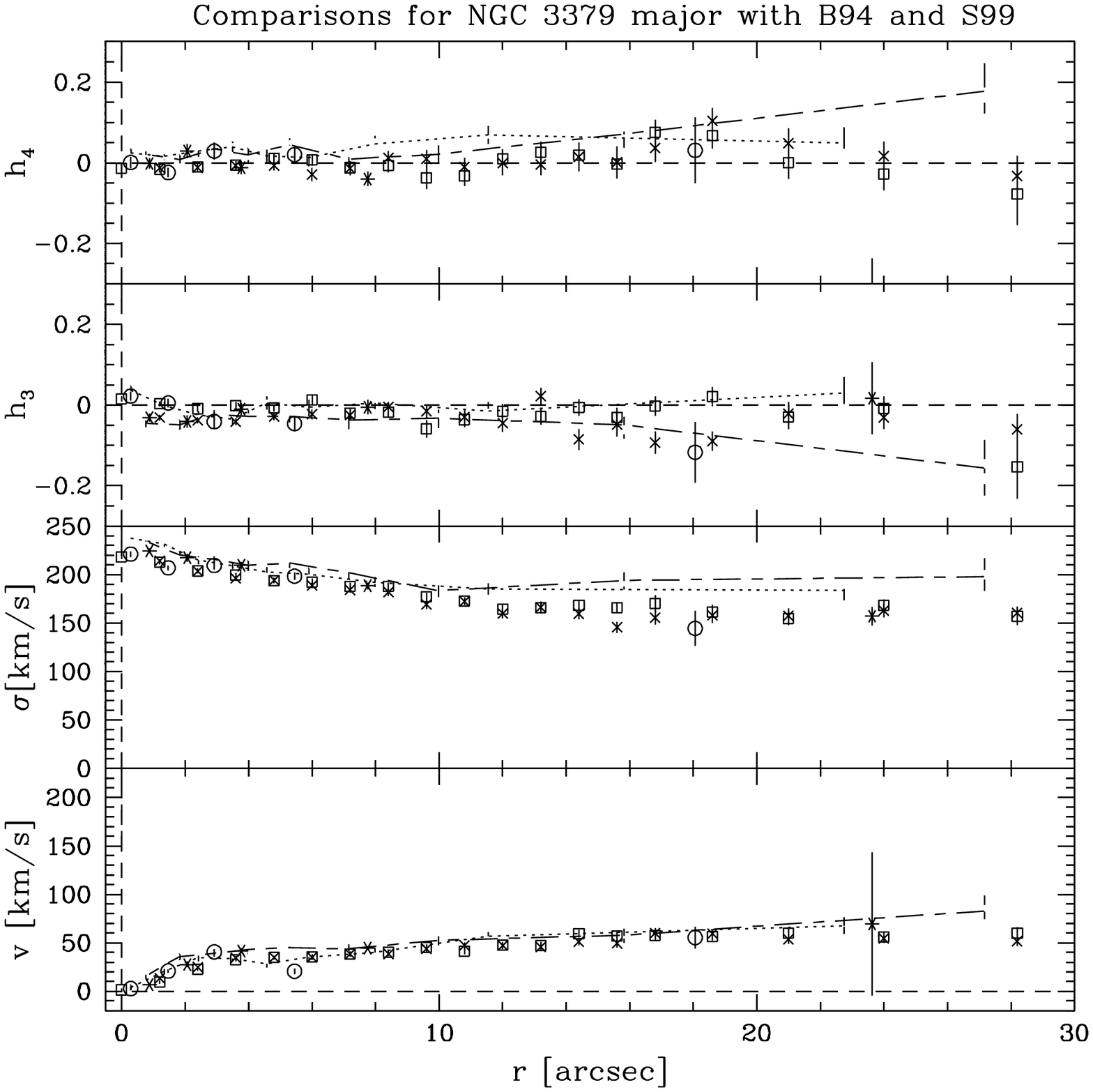,width=7.5cm}
\epsfig{file=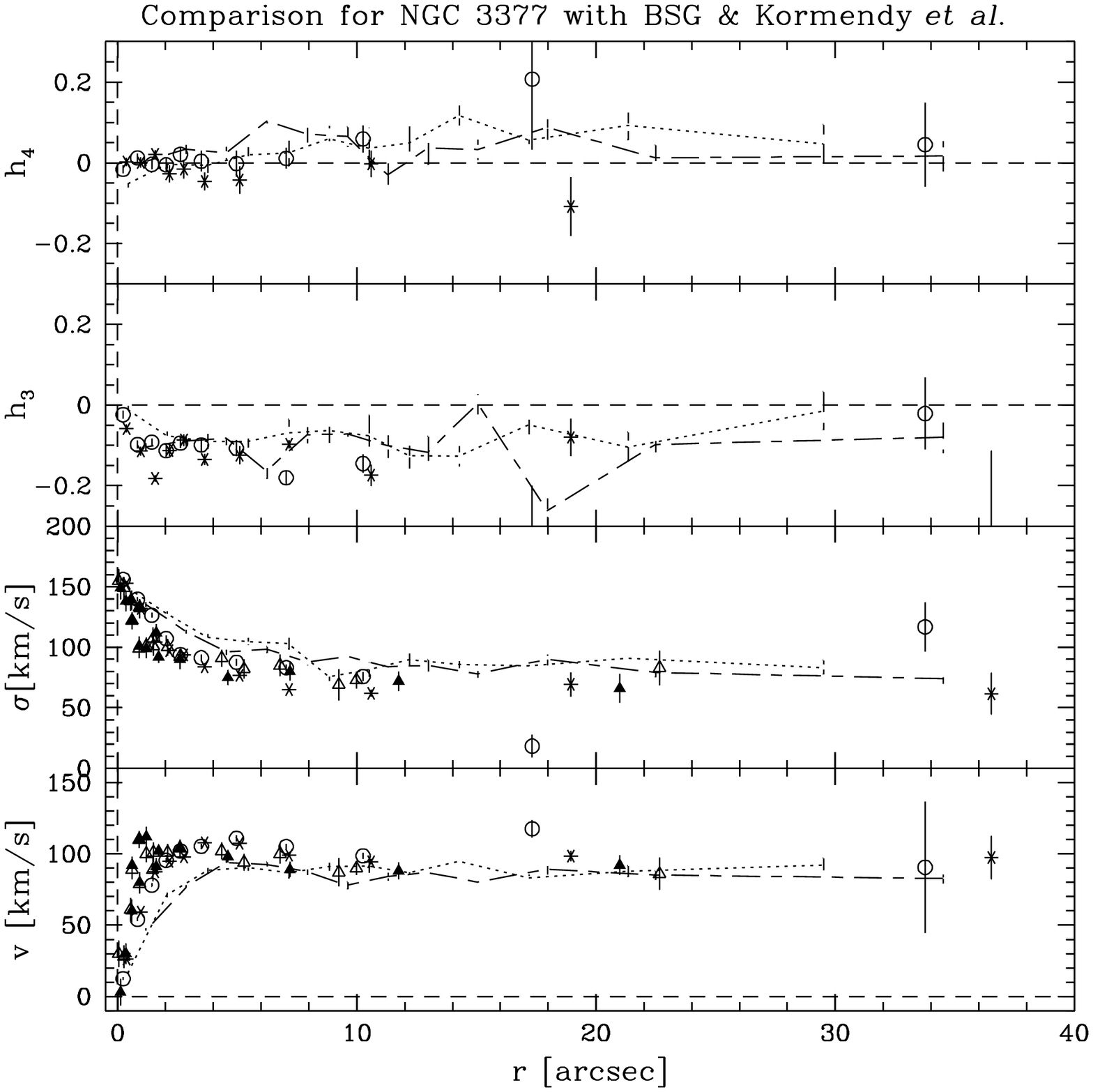,width=7.5cm}
\epsfig{file=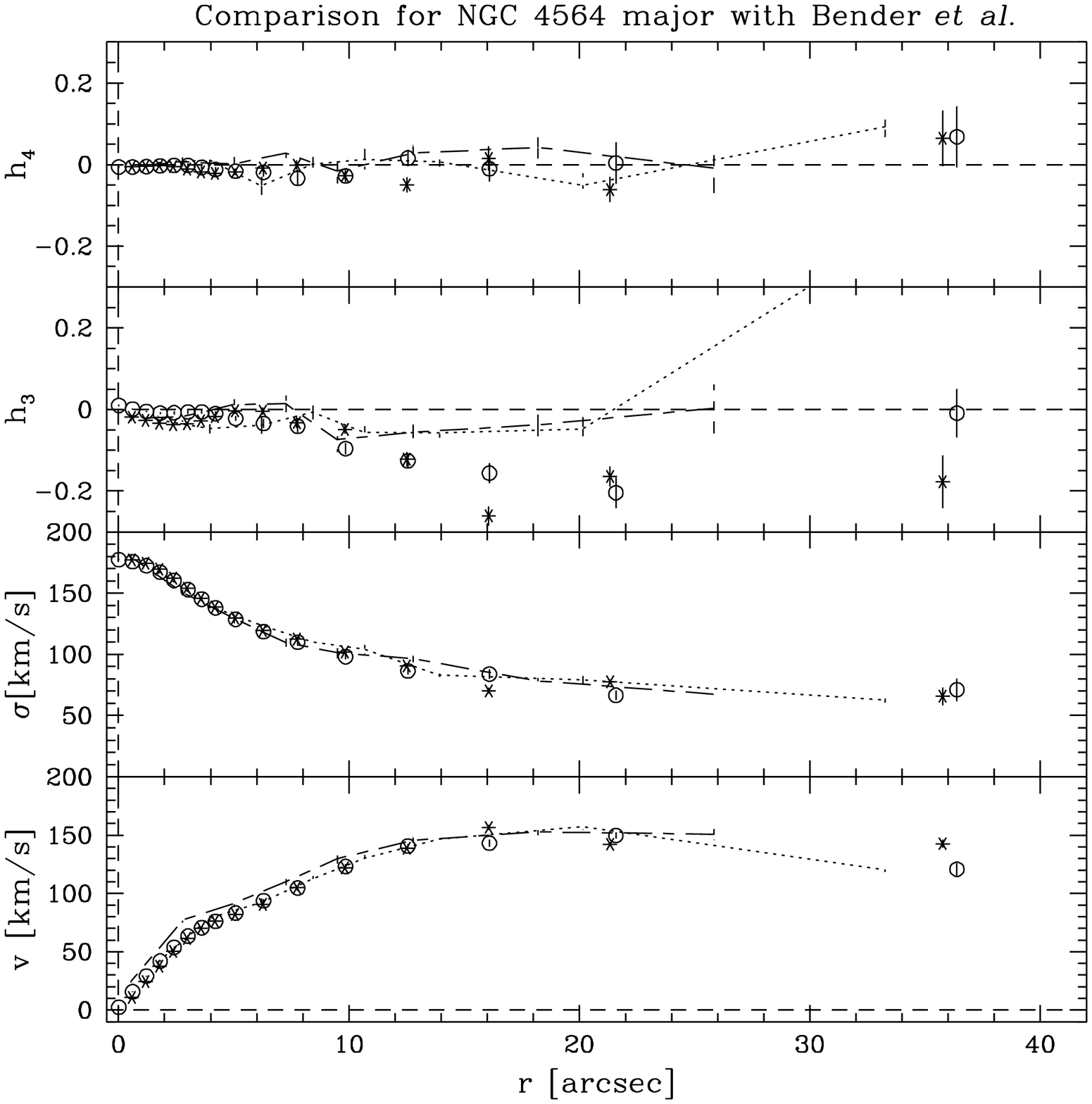,width=7.5cm}
\end{minipage} 
\caption[]{\label{cfBSG94K98} \small Comparison of measurements of rotation
  velocity, $\sigma$, $h_{3}$ and $h_{4}$ for galaxies in common with
  those studied by BSG94, SSH99 and K98. Measurements obtained in this
  study are indicated by open circles and asterisks for measurements
  in the direction and away from the direction of the PA of
  observation, respectively. BSG94 measurements are connected by dot
  and dashed lines for measurements in and away from the direction of
  the PA, respectively. SSH99 measurements are given by open and
  crossed square symbols for measurements in and away from direction
  of PA. K98 measurements are given by open and solid triangle symbols
  for measurements in and away from PA direction.}
\end{center}
\end{figure*}

Comparisons are made between measurements of v, $\sigma$, $h_{3}$ and
$h_{4}$ obtained here and measurements obtained previously by other
authors (BSG94, Kormendy et~al. 1998, hereafter K98, and SSH99).
Measurements have been obtained for the major axis of NGC\,3379,
NGC\,3377 and NGC\,4564 by BSG94, for the major axis of NGC\,3379 by
SSH99, and for rotation velocity and $\sigma$ of NGC\,3377 by K98. The
PAs of different observations for a given galaxy differed slightly,
but in all cases by less than 5\degr. Results of comparisons are given
in Figure \ref{cfBSG94K98}.

There is in general very good agreement between our measurements and
those of other studies. Disagreements can be attributed to the
different seeing conditions, instrumental set-ups and exposure
times. For NGC 3377 there is remarkably good agreement for the
higher-order terms $h_{3}$ and $h_{4}$. There are however differences
in the measurements of rotation and $\sigma$ at the galaxy centre. The
better spatial resolution of our own observations and particularly
those of K98 are able to resolve the steeply rising rotation curve;
the $\sigma$ measurements of K98 are also lower for these radii than
both our own and those of BSG94, reflecting their improved resolution
of the rotation curve. For NGC\,3379 our measurements of rotation,
$\sigma$, $h_{3}$ and $h_{4}$ agree well with the measurements of
BSG94 and SSH99. For NGC\,4564, there is disagreement between our
measured values of $h_{3}$ for $r \ga 12\arcsec$ and those of BSG94.

\end{document}